# First-in-human spinal cord tumor imaging with fast adaptive focus tracking robotic-OCT

Bin He[1,2#], Yuzhe Ying[3#], Yejiong Shi[1,2#], Zhe Meng[3], Zichen Yin[1,2], Zhengyu Chen[1,2], Zhangwei Hu[1,2], Ruizhi Xue[1,2], Linkai Jing[3], Yang Lu[3], Zhenxing Sun[3], Weitao Man[3], Youtu Wu[3], Dan Lei[3], Ning Zhang[4], Guihuai Wang[3*] and Ping Xue[1,2*]

[1]*State Key Laboratory of Low-dimensional Quantum Physics and Department of Physics, Tsinghua University*

[2]*Frontier Science Center for Quantum Information, Beijing, China*

[3]*Department of Neurosurgery, Beijing Tsinghua Changgung Hospital, School of Clinical Medicine and Institute of Precision Medicine, Tsinghua University, Beijing, 102218, China.*

[4]*Institute of Forensic Science, Ministry of Public Security, Beijing, 100038, China*

[#] *These authors contributed equally.*

* *youngneurosurgeon@163.com*

* *xuep@tsinghua.edu.cn*

## Abstract

Current surgical procedures for spinal cord tumors lack in vivo high-resolution, high-speed multifunctional imaging systems, posing challenges for precise tumor resection and intraoperative decision-making. This study introduces the **F**ast **A**daptive Fo**c**us **T**racking **R**obotic **O**ptical **C**oherence **T**omography (FACT-ROCT) system, designed to overcome these obstacles by providing real-time, artifact-free multifunctional imaging of spinal cord tumors during surgery. By integrating cross-scanning, rapid adaptive focal tracking and robotics, the system addresses motion artifacts and resolution degradation caused by intraoperative tissue movement, achieving wide-area, high-resolution imaging of tumors. We conducted intraoperative imaging on 21 patients, including 13 with spinal gliomas (grades I-IV) and 8 with other tumors or lesions. This study marks the first demonstration of OCT *in situ* imaging of human spinal cord tumors, providing micrometer-scale in vivo structural images and demonstrating FACT-ROCT's potential to differentiate various tumor types in real-time. Analysis of the attenuation coefficients of spinal gliomas revealed increased heterogeneity with higher malignancy grades. So, we proposed the standard deviation of the attenuation coefficient as a physical marker, achieving over 90% accuracy in distinguishing high- from low-grade gliomas intraoperatively at a threshold of 0.75 $mm^{-1}$. FACT-ROCT even enabled extensive in vivo microvascular imaging of spinal cord tumors, covering 70 mm * 13 mm * 10 mm within 2 minutes. Quantitative vascular tortuosity comparisons confirmed greater tortuosity in higher-grade tumors. The ability to perform extensive vascular imaging and real-time tumor grading during surgery provides critical information for surgical strategy, such as minimizing intraoperative bleeding and optimizing tumor resection while preserving functional tissue. Overall, FACT-ROCT offers high-resolution, high-speed, and comprehensive intraoperative imaging of spinal cord tumor structure and vasculature, with the potential to improve surgical outcomes by aiding surgeons in making more informed decisions.

## 1. Introduction

The spinal cord, along with the brain, forms the central nervous system, essential for the body's functioning. The spinal cord is a quasi-cylindrical structure, about 10 to 14 millimeters in diameter, running from the brain down through the vertebral column [1]. Acting like an optical fiber cable, it transmits signals between the brain and the rest of the body, coordinating motor functions, sensory information, and autonomic responses. Because of its critical role, any abnormal growths, especially tumors, in or around the spinal cord can cause severe neurological impairments. These impairments can result in pain, loss of motor function, sensory deficits, and autonomic dysfunction, significantly affecting a patient's quality of life [2].

Surgery is the primary treatment for spinal cord tumors, aiming to remove as much of the tumor as possible while preserving normal neural structures [3]. The spinal cord's limited space and critical functions make this a particularly challenging task, necessitating high-resolution imaging to accurately identify vital normal structures, such as nerve roots, fasciculus or grey matter and avoid damaging them during surgery. Current imaging technologies like MRI [4], CT [5], and ultrasound [6], while beneficial, often fall short in providing the necessary resolution and real-time capabilities needed for optimal intraoperative guidance. These modalities may struggle with clearly delineating the tumor from surrounding critical normal structures and unable to differentiate tumors between types, which is crucial for effective surgical planning and execution [7]. Moreover, accurate tumor grading during surgery is also essential for determining the appropriate surgical strategy. High-grade tumors, such as advanced spinal gliomas, typically require more extensive resection due to their aggressive nature [8]. However, the extent of resection must be carefully balanced against the risk of damaging critical neural pathways. Accurate grading helps optimize patient outcomes by tailoring the surgical approach to the tumor's aggressiveness. Currently, intraoperative tumor grading is primarily achieved through frozen section pathology [9]. While this method can provide reasonably accurate results (about 70%), it also has limitations, such as being time-consuming, typically taking around one hour in clinical practice, and the potential for sampling errors, as the small tissue sample may not be representative of the entire tumor [10].

Additionally, functional vascular imaging is crucial in spinal cord tumor surgery, as detailed vascular maps allow surgeons to avoid blood vessels during resection, minimizing intraoperative bleeding and improving surgical safety. Moreover, the significant differences in the vascular structures of tumors compared to surrounding healthy tissue are important markers of tumor characteristics [11]. International diagnostic standards, such as the WHO [12], Kernohan system [13], and St. Anne/Mayo grading system [14], consider the formation of new blood vessels within the tumor as a key indicator of its malignancy. Understanding the tumor's vascular characteristics also aids in assessing its aggressiveness and planning further treatments. In terms of vascular imaging during spinal cord tumor surgery, techniques such as magnetic resonance angiography (MRA) [15], computed tomography angiography (CTA) [16], and fluorescence angiography [17] are commonly used. MRA and CTA can provide vascular structures with low resolution and lack real-time application during surgery. While fluorescence angiography, though providing real-time visualization, is limited by its lack of tomographic capabilities and the need for dye injection. Overall, there remains a significant gap in the availability of a high-speed, in vivo imaging modality that can effectively differentiate between types and grades of spinal cord tumors and perform in situ vascular imaging during surgery.

Optical Coherence Tomography (OCT) is an emerging non-invasive, high-resolution imaging modality that provides three-dimensional, label-free visualization of biological tissues, primarily within 1 to 2 mm depths [18]. OCT has been widely used in clinical applications, particularly in ophthalmology [19] and cardiovascular imaging [20]. In recent years, OCT has also been applied to brain tumor imaging [21]. Initial studies focused on ex vivo imaging of human brain tumors in both 2D or 3D formats [22]. Kut et.al applied OCT to glioma resection surgeries, [23] performing in situ imaging of fresh human brain tumor tissues, demonstrating that OCT could differentiate tumor tissue from normal tissue based on attenuation coefficients. However, these studies primarily involved ex vivo tissues, which differ significantly from in vivo tissues in aspects such as blood perfusion. Almasian et al. extended OCT imaging to in vivo human brain tumors [24], but their handheld probe had stability issues, preventing effective vascular imaging and limiting the imaging range. Kuppler et al. integrated OCT systems with surgical microscopes [25,26], achieving in vivo OCT brain tumor images. However, their system had lower resolution and lacked functional vascular imaging. Vakoc et al. applied OCT angiography imaging to mouse brain tumors [27], showing distinctions between tumor and normal tissue vasculature. Nonetheless, they did not extend their research to human brain tumors. Ramakonar et al. [28] further explored in vivo vascular imaging with an OCT probe, yet their findings were confined to B-scan images and lacked volumetric data, thus limiting clinical translation. Overall, OCT still faces several challenges in clinical applications. Intraoperative OCT imaging demands a broader imaging range to cover entire tumor areas. Additionally, to achieve multifunctional in vivo OCT imaging, the imaging system must be highly stable, overcoming motion artifacts caused by living tissues such as heartbeats. Moreover, while OCT has been primarily used for intraoperative imaging of human brain tumors, its application in spinal cord tumor surgery remains unexplored, representing a significant gap in current research.

To address these challenges, we introduce a **F**ast **A**daptive Fo**c**us **T**racking **R**obotic **O**ptical **C**oherence **T**omography (FACT-ROCT) system for intraoperative spinal cord tumor imaging. By mounting the OCT probe on a force-controlled 7 degrees-of-freedom (DOF) robotic arm, we can achieve continuous large-area automatic imaging, covering the entire spinal cord tumor region safely. To overcome deformation and lateral resolution degradation caused by motion during surgery, we use rapid cross-sectional B-scans to acquire surface topology and an electrically tunable lens (ETL) for focus tracking. By utilizing the surface depth information obtained from one B-scan to drive the focus of the subsequent B-scan, we achieve a focus tracking speed of ~10 milliseconds. This enables high-resolution three-dimensional imaging of various shapes, mitigating resolution degradation due to object motion.

Using the FACT-ROCT system, we conducted intraoperative OCT imaging on 21 patients with spinal cord tumors, including 13 with spinal gliomas (grades I to IV) and 8 with other tumors or lesions such as vascular reticulocytomas and teratomas. This study marks the first demonstration of OCT *in situ* imaging of human spinal cord tumors, providing micrometer-scale in vivo structural images of various tumor types. Our findings reveal the potential of FACT-ROCT to differentiate between high- and low-grade gliomas with over 90% accuracy in real-time tumor grading based on attenuation coefficient heterogeneity. Additionally, the FACT-ROCT system enabled extensive vascular imaging, covering an area of 70 mm * 13 mm * 10mm in under 2 minutes. Detailed vascular maps provided critical information for surgical planning, with quantitative comparisons confirming greater vascular tortuosity in higher-grade tumors. Our study demonstrates the translational potential and practicality of FACT-ROCT in spinal cord tumor surgery, paving the way for its integration into clinical practice.

## 2. Result

### 2.1 FACT-ROCT imaging approach

We conducted a first-in-human study using the FACT-ROCT system on patients undergoing spinal cord tumor resection surgeries. The FACT-ROCT system employs a probe mounted on a 7-DOF robotic arm (Fig. 1A and Fig.1 C), which features force control capabilities. This robotic arm supports near zero-force dragging and virtual wall control, ensuring optimal safety during intraoperative imaging. The FACT-ROCT system is based on a customized swept-source OCT system that utilizes a high-speed (200kHz) near-infrared (1310nm) laser, providing a long coherence length and high axial resolution of ~10 μm within the tissue. This system achieves three-dimensional structural and vascular imaging of the tissue by scanning with fast and slow-axis galvo mirrors, and collecting the backscattered light from the different layers of the tissue, as detailed in the Materials and Methods section.

Prior to imaging, the surgeon uses MRI images to determine the approximate location of the tumor. The surgical procedure involves making an incision through the skin and muscles, removing the lamina, and opening the dura mater to expose the spine and tumor (Fig. 1B). These steps are standard preoperative procedures for spinal cord tumor resection surgery. For safety, the surgeon firstly uses real-time OCT cross-sectional images to manually guide the probe to the leading edge of the spinal cord tumor with the robotic arm. Once the initial OCT volume is obtained, the robotic arm automatically moves to the next imaging position based on the previous OCT volume (See methods for more details), enabling extensive imaging of the entire spinal cord tumor region (Fig. 1C). The FACT-ROCT system can acquire 280 cross-sectional images per second and displaying them in real-time. It enables volumetric structural imaging in approximately 2.5 seconds and volumetric vascular imaging within 10 seconds. The single scan range of the system reaches 13 mm x 13 mm x 5 mm, covering the diameter of the spinal cord and allowing for extensive lesion coverage through the movement of the robotic arm.

During intraoperative OCT imaging, tissue movement caused by the patient's heartbeat is inevitable, leading to two significant issues. Firstly, it causes severe deformation of the OCT volume, manifesting as a hump-like distortion, as shown in Fig. 1D. To solve this issue, we perform a fast cross-sectional scan using the original slow-axis galvo mirrors before each OCT volumetric imaging session. This approach captures the object's original contour along the slow axis, as indicated by the red line in Fig. 1D, allowing for the correction of the object's original three-dimensional shape with minimal added imaging time.

Additionally, heartbeat-induced jitter can also cause frequent defocusing of the imaging plane. To solve this, we propose a novel rapid adaptive focus tracking technique to maintain focus during imaging. It involves a feedback loop for continuous focus refinement without needing prior knowledge of the object's shape. Specifically, an ETL is integrated into the OCT probe. The depth information obtained from the previous B-scan can drive the ETL to achieve the accurate focus of the next B-scan, thereby enabling almost all B-scans to be focused (Fig.1 E), achieving focus tracking speeds of ~10 milliseconds. The capability of fast focus tracking makes the FACT-ROCT suitable for tissues with curved and tilted surfaces in addition to the tissue movement, it could maintain consistent focus and image quality across various anatomical structures (Fig.1 F). Unlike traditional ETL-based OCT [29-31], which requires multiple volumetric scans or pre-scans to maintain high resolution for complex curved objects, thereby increasing imaging time, FACT-ROCT can achieve high resolution imaging of the entire curved object with a single volumetric scan based on real-time focus tracking. Additionally, this method does not suffer from sensitivity loss like special beams such as needle beams [32] or Bessel beams [33], and it can be applied to any OCT system, demonstrating great versatility.

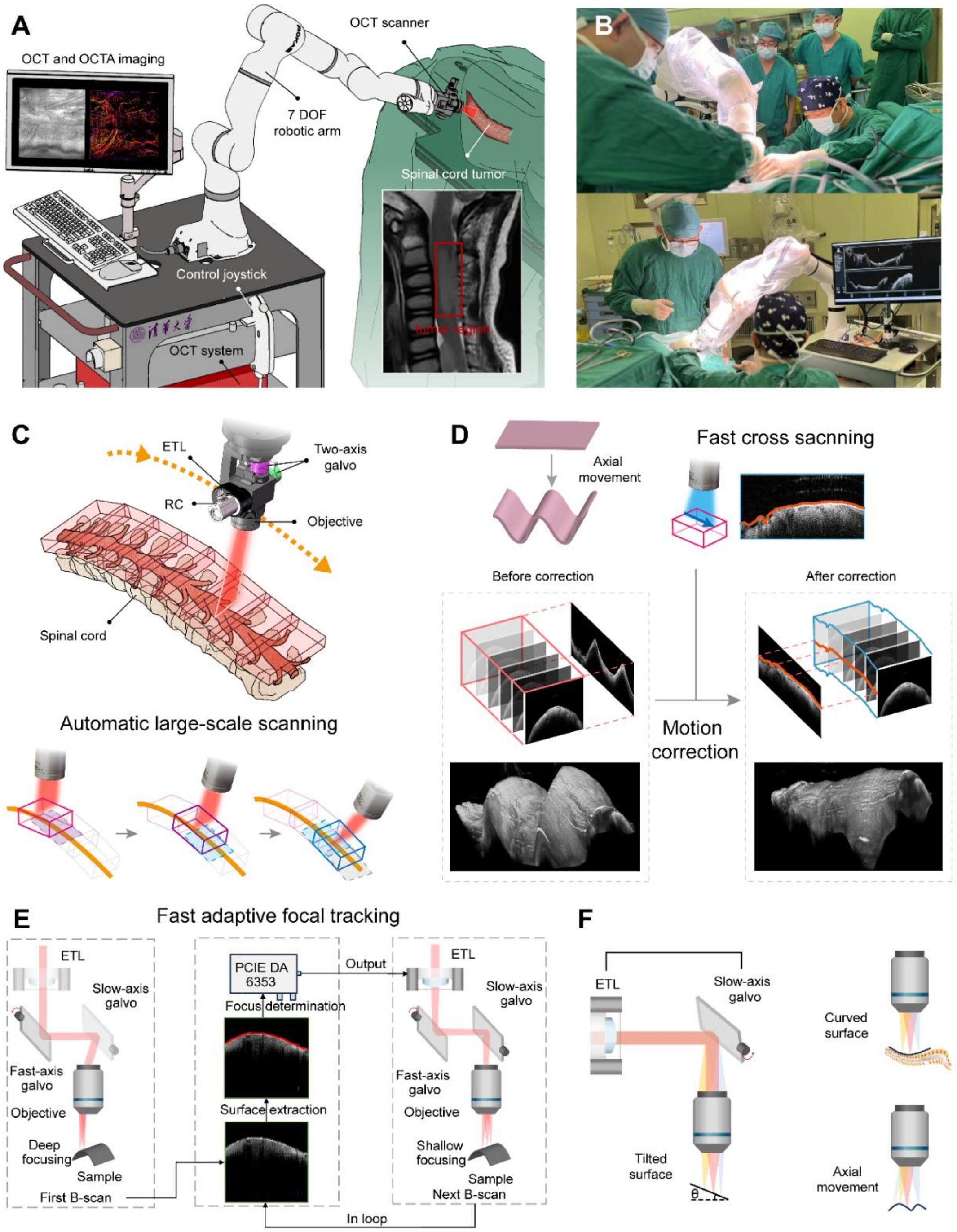

**Fig. 1. Schematic and Application of the FACT-ROCT System for Intraoperative Spinal Cord Tumor Imaging.** (A) Overview of the FACT-ROCT system setup, which includes an OCT scanner mounted on a 7-DOF robotic arm. The system performs both OCT and OCT angiography imaging of spinal cord tumors. The inset shows a spinal cord tumor in an MRI scan, indicating the region of interest. (B) Photographs demonstrating the use of the FACT-ROCT system during spinal cord tumor surgery. (C) Schematic of the automatic large-scale scanning mechanism employed by the FACT-ROCT system. *Abbreviations:* RC, reflective collimator; ETL, electrically tunable lens. (D) Illustration of the fast cross-sectional scanning and motion correction process. (E) Diagram of the fast adaptive focal tracking method. The system uses surface depth information obtained from one B-scan to drive the focus of the subsequent B-scan. This process involves a feedback loop for continuous focus refinement. (F) Depiction of the FACT-ROCT's ability to adapt to curved and tilted surfaces, as well as axial movement, maintaining consistent focus and image quality across various anatomical structures.

## 2.2 Focus Tracking Performance of the FACT-ROCT System

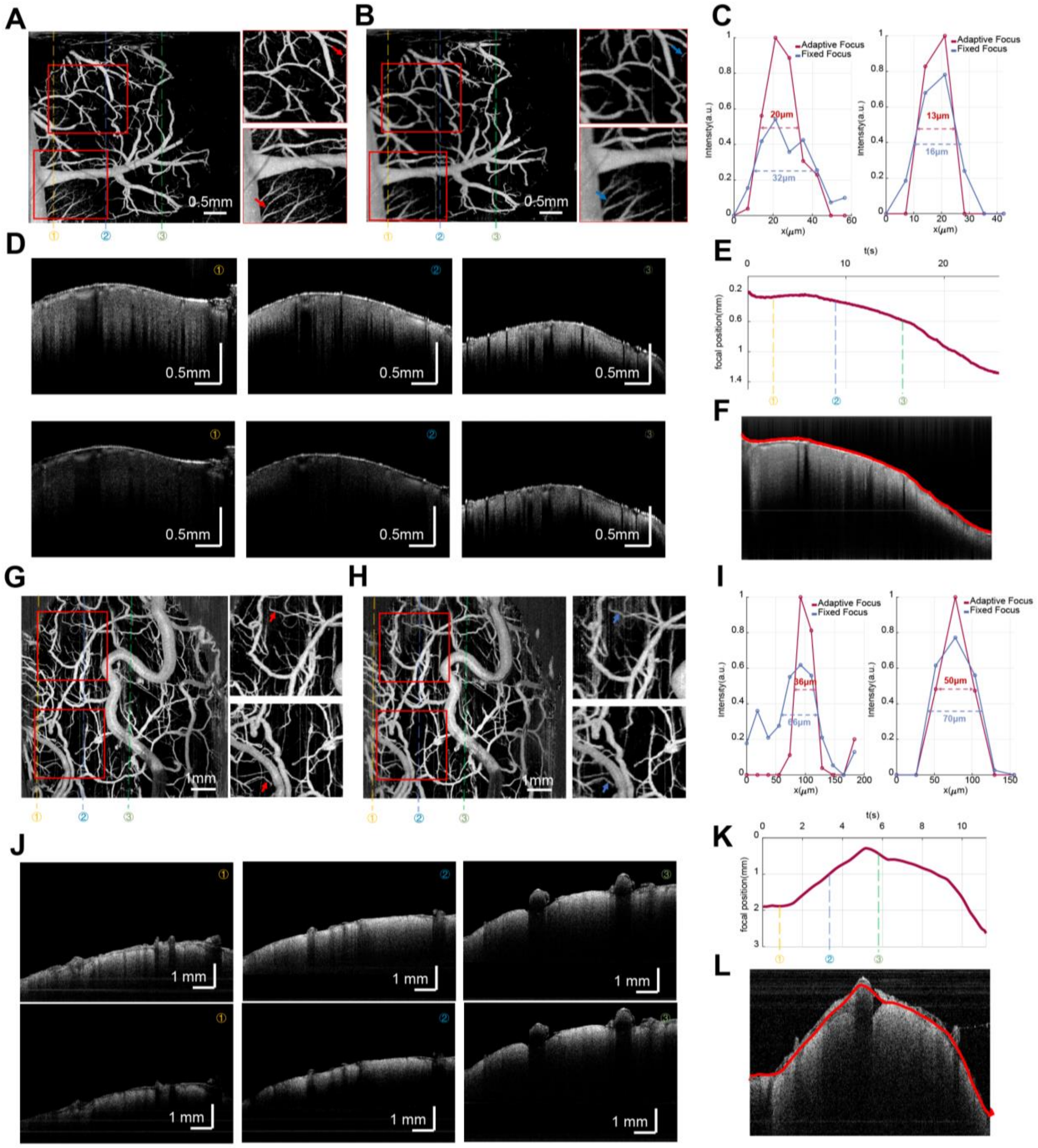

**Fig. 2. Evaluation of the focal tracking ability of the FACT-ROCT System.** (A) - (F) represent tests with high lateral resolution (8 μm). (G) - (L) represent tests with low lateral resolution (25 μm). OCT angiography images of a mouse brain and a human spinal cord tumor are shown using adaptive focus (A, G) and fixed focus (B, H) techniques, the entire region of panel A shows clear vascular imaging, while in panel B, the left half is blurry. Similarly, in panel G, the entire region is clear, while in panel H, the vessels are less distinct. Insets highlight regions of interest. (C, I) Quantitative analysis of vessel diameter for adaptive focus (red line) versus fixed focus (blue line). The vessels selected are indicated by arrows in the insets of panels A, B, G and H (D, J) OCT B-scan images at different positions corresponding to the dashed lines in panels A, B, G, and H, showing images obtained with adaptive focus (top row) and fixed focus (bottom row). (E, K) Focal position correction curves over time used for adaptive focusing, corresponding to the surface profiles in the slow axis direction OCT B-scan (F, L).

We further evaluate the focus tracking ability of the FACT-ROCT on tissues with various shapes. The evaluations are conducted using both high lateral resolution (8 μm) and low lateral resolution (25 μm) scan objectives to demonstrate its versatility, as shown in Fig. 2. For the high lateral resolution tests, we perform OCT imaging on a mouse brain. The significant curvature of the mouse brain cortex, with over 1 mm depth variation limits traditional fixed-focus techniques to achieving focus in only a small region, adversely affecting both OCT structural and angiographic imaging. In contrast, by utilizing the continuous focus refinement feedback of the FACT-ROCT, we can achieve uniform focusing across all B-scans by applying the focal curve shown in Fig. 2E to the ETL. The results demonstrate that the adaptive focus technique (Fig. 2A) provides clear vascular imaging across the entire region, whereas the fixed focus technique (Fig. 2B) results in blurred vessels on the left side. Quantitative analysis of vessel diameter (Fig. 2C) also confirms this, showing a 60% improvement in resolution in some areas with the adaptive focus technique compared to the fixed-focus technique. The OCT structural images further highlight this difference. Our FACT-ROCT maintains focus across all B-scans, resulting in consistent and better image signal-to-noise ratio (SNR) and resolution. In contrast, traditional OCT achieves focus only in the final few B-scans, with the remaining B-scans appearing blurred and with low SNR (Figure 2D).

We also compare the performance of FACT-ROCT with traditional OCT in imaging human spinal cord tumors intraoperatively, based on low lateral resolution (25 μm) but with a larger field of view of 13mm x 13mm. As shown in Fig. 2G-L, the results involve a patient diagnosed with a WHO Grade II Ependymoma post-surgery. Due to the significant curvature of the imaging target, with nearly 2.5mm depth variation (Fig. 2K), traditional OCT results show blurred vascular imaging on both sides, with partial vascular absence on the right side (Fig. 2H). In contrast, by employing focus tracking, the FACT-ROCT is able to achieve high resolution vascular imaging across the entire field of view (Fig. 2G), demonstrating that the tumor's vasculature maintains a vertical orientation with a slight increase in tortuosity. Further quantitative analysis also underscores the advantages of our method in terms of vascular resolution and SNR. Moreover, structural images reveal that in traditional OCT, only the B-scans located at the central part of the object remain in focus, with others being out of focus, whereas FACT-ROCT maintains high-resolution status across all B-scans with a significantly improved SNR. Additionally, the FACT-ROCT system can achieve focus tracking even in the presence of patient's longitudinal jitter up to 3 mm during intraoperative imaging, maintaining consistent lateral resolution and high SNR (SFig.2).

## 2.3 Intraoperative Imaging of Spinal Cord Tumors with FACT-ROCT

We have utilized FACT-ROCT to perform intraoperative spinal tumor imaging on 21 patients, serving as a complementary imaging modality alongside the surgical microscope to assess tumor characteristics. The patient cohort, detailed in Table S1, primarily included spinal cord gliomas such as ependymomas and diffuse midline gliomas (DMGs), as well as other tumors like hemangiomas and teratomas. For each patient, multiple scans were performed using the robotic arm to achieve complete coverage of the tumor area, with all imaging mostly completed within 2 minutes. Data was mostly collected both before and after tumor resection. Additionally, given that OCT is a non-contact, label-free, and non-destructive imaging technique, FACT-ROCT did not cause any imaging-related side effects, and no adverse events associated with OCT use were observed, underscoring its safety and tolerability in clinical practice.

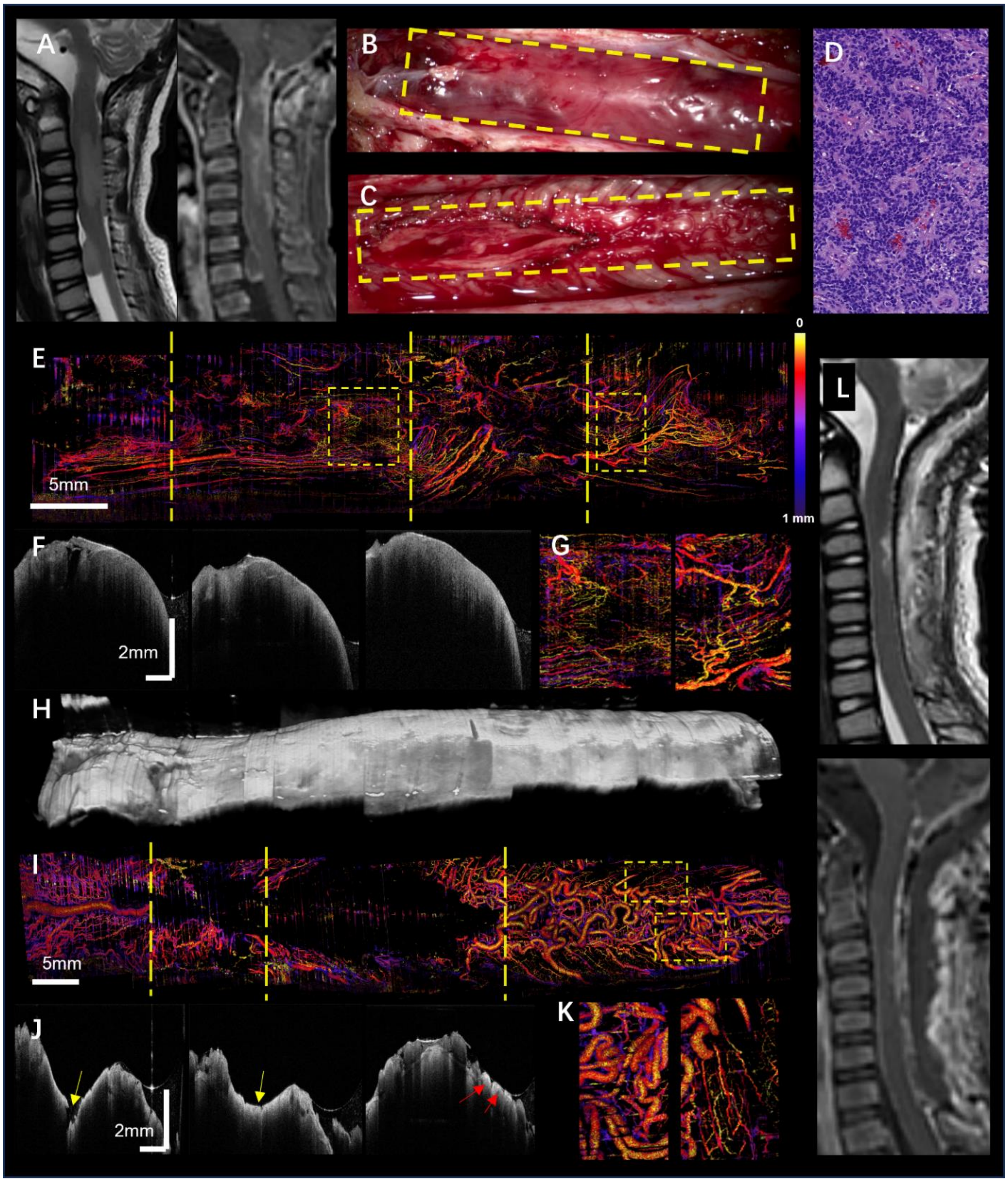

**Fig.3. A case of recurrent ependymoma with FACT-ROCT:** (A) Preoperative MRI shows the tumor in the cervical spinal canal, causing fusiform swelling and surface breakthrough. Contrast-enhanced MRI reveals heterogeneous signals. (B) Intraoperative microscopic image of the tumor, located on the spinal cord surface with a rich blood supply, though surface vasculature appears sparse. (C) Post-tumor resection, showing the surgical cavity. (D) Postoperative pathology shows moderately dense tumor cells with moderate atypia, forming rosettes and perivascular pseudorosettes, characteristic of WHO Grade III ependymoma. (E) OCT angiography shows dense and irregular tumor vasculature, with large surface vessels corresponding to panel B. (F) Selected OCT B-scan images from the tumor's rostral to caudal ends, displaying heterogeneous scattering signals. (G) Magnified vascular image from the yellow dashed box in panel E, showing dense and heterogeneous tumor vasculature. (H) 3D OCT structural image of the entire tumor region shows its three-dimensional morphology. (I) Post-resection OCTA image reveals no vascular signals within the surgical cavity, with surface vessels corresponding to panel C. (J) Selected post-resection OCT B-scan images shows uniform optical signals at the tumor bed (yellow arrows). Blood vessels and nerve roots are observed on the caudal surface (red arrows) (K) Magnified vascular image from the yellow dashed box in panel E. (L) Postoperative MRI shows no residual tumor.

Here, we present a detailed case to demonstrate the utility of OCT and OCT angiography (OCTA) in intraoperative spinal tumor imaging. The patient, a 4-year-old male, had undergone intracranial ependymoma resection three years ago. Preoperative MRI (Fig.3 A) revealed two intramedullary lesions in the cervical spinal canal, the larger measuring approximately 59 x 10 mm, with significant and heterogeneous signals upon contrast-enhanced MRI. During the surgery, after incising the dura mater and suspending it, the tumor was fully exposed. The tumor extended from the fourth ventricular outlet to the C7 vertebral level, appearing soft, grayish-red, and lobulated, suggesting a rich blood supply. However, the surgical microscope could hardly visualize the microvasculature (Fig.3 B). After partial tumor resection, it was observed that parts of the tumor had invaded the spinal cord parenchyma at the C3-C5 vertebral levels. Complete tumor resection was achieved, and the surgical field at the C3-C7 vertebral levels post-resection is shown (Fig.3 C). Postoperative pathological examination (Fig.3 D) revealed highly dense tumor cells with moderate atypia, forming rosettes and perivascular pseudorosettes, consistent with WHO Grade III ependymoma.

Following tumor exposure, we employed FACT-ROCT to image the entire tumor region. The OCT angiography image (Fig.3 E) revealed a rich and heterogeneous blood supply within the tumor. The magnified region of interest (Fig.3 G) further demonstrated the high density and tortuosity of the tumor, consistent with the characteristics of WHO Grade III ependymoma microvascular proliferation. This characteristic of tumor vessels differs significantly from the vascular features of WHO Grade II ependymoma tissue presented in the previous section, suggesting that vascular characteristics could serve as markers for tumor grading. Additionally, the OCT angiography results of this patient revealed an uneven distribution of vascular density within the tumor region, with the left side showing a lower vascular density compared to the right side. This feature could guide strategies for the safe tumor resection during surgery. OCT B-scan structural images (Fig.3 F) showed heterogeneous scattering signals within the tumor, indicative of significant tumor heterogeneity, consistent with high-grade malignancy. And the 3D OCT structural images (Fig.3 H) depicted the tumor's three-dimensional morphology. Post-resection OCT angiography images (Fig.3 I, K) showed vascular occlusion and disrupted blood supply in the tumor-invaded areas due to bipolar coagulation, while surface vasculature was preserved in non-invaded spinal cord regions. Post-resection OCT structural images (Fig.3 J) demonstrated uniform optical signals within the spinal cord parenchyma at the tumor bed (indicated by yellow arrows). Some blood vessels and nerve roots were observed on the caudal surface of the surgical area (indicated by red arrows). Follow-up MRI (Fig.3 L) confirmed the absence of residual tumor. This case illustrates the role of FACT-ROCT in enhancing intraoperative visualization of tumor vasculature and structure, facilitating precise and safe tumor resection, and ensuring postoperative assessment. Additional cases, such as diffuse midline glioma and subependymal tumor, are presented in the supplemental materials in SFig.3 and SFig.4.

## 2.4 Comparative Imaging of Normal and Tumor Tissues Using FACT-ROCT

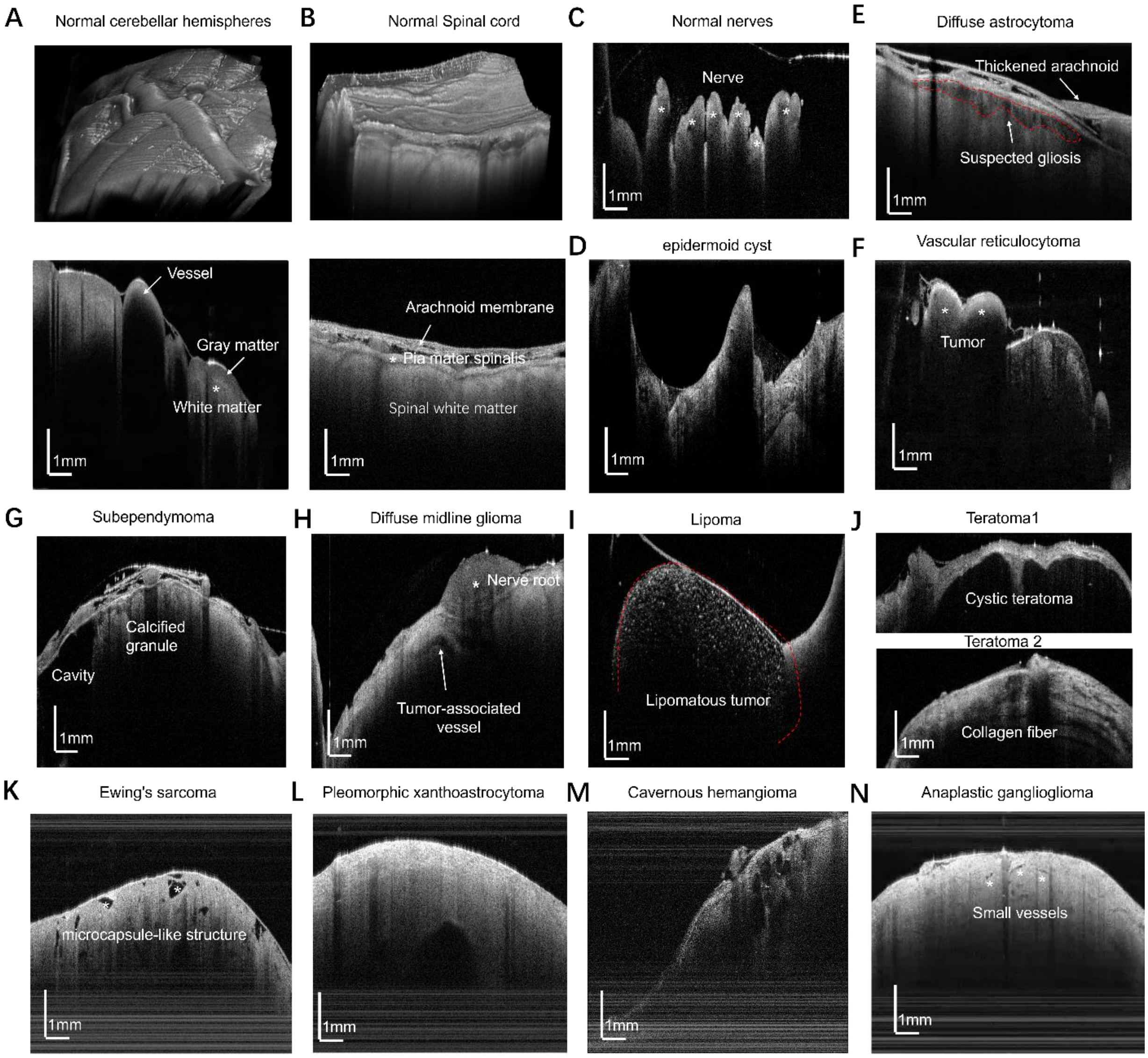

**Fig.4. Images of normal tissues and different spinal cord tumors obtained by FACT-ROCT :** (A) 3D OCT structural image of the normal human cerebellar hemisphere and one OCT B-scan image, highlighting a clear boundary between the upper gray matter and lower white matter. (B) 3D OCT structural image of the normal human spinal cord and one OCT B-scan cross-section, showing a uniform and continuous dark band between the spinal cord white matter and pia mater, with small blood vessels visible in the subarachnoid space. (C) OCT image of normal nerve roots, with nerve roots float in the cerebrospinal fluid. (D) OCT image of an epidermoid cyst, with cyst contents below the cyst wall showing mixed scattering signals. (E) Image of diffuse astrocytoma with suspected gliosis visible. (F) OCT image of a vascular reticulocytoma with an irregular shape located at the normal medullary-cervical junction. (G) Image of a subependymal tumor showing features such as cavities and calcified granules. (H) OCT image of a DMG showing a nerve root located on the tumor surface and tumor-associated abnormal vessels. (I) OCT image of a lipoma with hyperintense fat granules. (J) OCT images of teratomas, with one image showing cyst and the other showing collagen fibers. (K) OCT images of ewing's sarcoma with the feature of microcapsule-like structure. (L) OCT images of pleomorphic xanthoastrocytoma with an irregular cystic change in the deep part of the tumor. (M) OCT images of cavernous hemangioma showing the feature of Spongiform changes. (N) OCT images of Anaplastic ganglioglioma with small blood vessels passing through the tumor.

In this section, FACT-ROCT was utilized to obtain detailed imaging of normal tissues and various spinal cord tumors. The images provided critical insights into the structural differences between these tissues, suggesting that FACT-ROCT has the potential in distinguishing different types of spinal cord tumors. For instance, we imaged a normal human cerebellar hemisphere and spinal cord individually. The 3D OCT structural image of the cerebellar hemisphere (Fig. 4 A) clearly highlighted the boundary between the upper gray matter and the lower white matter. In contrast, the 3D OCT image of the normal human spinal cord (Fig. 4 B) shows a different structure, with the white matter located above the gray matter. Additionally, there was a uniform and continuous dark band between the spinal cord white matter and pia mater, with small blood vessels visible in the subarachnoid space. The OCT image of normal nerve roots floating in the cerebrospinal fluid (Fig. 4 C) further demonstrated that FACT-ROCT was capable of identifying vital structures that were critical during surgery.

Moving to pathological tissues, OCT images provide some insights into different tumor types. The OCT image of epidermoid cyst (Fig.4 D) shows that the cyst contents are below the cyst wall and exhibit mixed scattering signals, indicating various tissue compositions. In the case of diffuse astrocytoma (Fig. 4E), suspected gliosis is evident (marked with a red dashed box). The vascular reticulocytoma (Fig. 4F) is seen as an irregularly shaped mass at the bulbar cervical medullary junction. The OCT image of subependymal tumor (Fig. 4G) reveals typical characteristic features such as cavities and calcified granules. Additionally, an OCT image of a DMG (Fig. 4H) shows a nerve root located on the tumor surface along with tumor-associated abnormal vessels, emphasizing the complex vascular structures that can be visualized. A lipoma with hyperintense fat granules is clearly illustrated in OCT image (Fig.4 I), while Figures 4J present OCT images of teratomas, one showing a cyst and the other displaying collagen fibers. Besides, Fig.4K-N show OCT images of Ewing's sarcoma, Pleomorphic xanthoastrocytoma, Cavernous hemangioma, Anaplastic ganglioglioma with their own unique features. These findings underscore the potential of FACT-ROCT in providing high-resolution images that differentiate between different types of spinal cord tumors, thereby facilitating effective intraoperative and postoperative assessments.

## 2.5 Analysis of OCT Structural Images for Spinal Cord Gliomas with Different Grades

We further utilized FACT-ROCT for intraoperative imaging of spinal cord gliomas of various grades, including two patients with Grade I, Four with Grade II, three with Grade III, and four with Grade IV tumors. We analyzed the in vivo OCT structural images obtained during these procedures. To ensure experimental accuracy, the imaged areas were selected based on MRI indications of tumor parenchyma, with tumor grades confirmed by postoperative pathological diagnosis. The detailed 3D OCT structural images, en face images, and cross-sectional images of an Astrocytoma (Grade III) (Fig. 5A) demonstrate the capability of FACT-ROCT to capture a broad tumor region, encompassing an area of 60mm x 13mm x 10mm. Additionally, the cross-sectional OCT structural images of tumors revealed that higher-grade tumors exhibit more uneven scattering signals. The typical OCT B-scan images (Figs. 5B-E) further depicted differences in tissue structure or composition among spinal cord gliomas of different grades.

To quantitatively analyze the differences in OCT images between grades, we developed a method to obtain the optical attenuation coefficient (OAC) of the tumors, eliminating depth-dependent effects of the beam profiles. Since FACT-ROCT consistently ensures the focus remains at a fixed position within the sample, we can obtain reliable OAC measurements across measurements (See methods for more details). Unlike previous studies on ex vivo human brain gliomas, we found significant heterogeneity in the OACs of in vivo spinal cord gliomas, with overlapping OAC values across different grades, making it difficult to distinguish tumor grades solely based on OAC. However, we observed notable differences in the variance of the OAC within a single B-scan across different tumor grades, which aligns with the understanding that tumor heterogeneity increases with higher grades. The box plots (Fig. 5F) highlight the distribution of the standard variance of OAC, showing a clear increase in the standard deviation of OAC with higher tumor grades. This suggests that the optical properties of gliomas become more variable as the grade increases, reflecting greater heterogeneity within the tumor tissues. Furthermore, the comparison of the standard variance of OAC between low-grade (Grade I and II) and high-grade (Grade III and IV) gliomas (Fig. 5G) underscores the significant differences in optical characteristics between these two categories. The sensitivity analysis (Fig. 5H) reveals that a threshold value of 0.75 mm$^{-1}$ for the standard variance of OAC can distinguish between low-grade and high-grade gliomas with over 90% accuracy. This high level of accuracy demonstrates the potential of FACT-ROCT in providing real-time tumor grading during surgery, which is crucial for determining the appropriate surgical strategy for spinal cord gliomas.

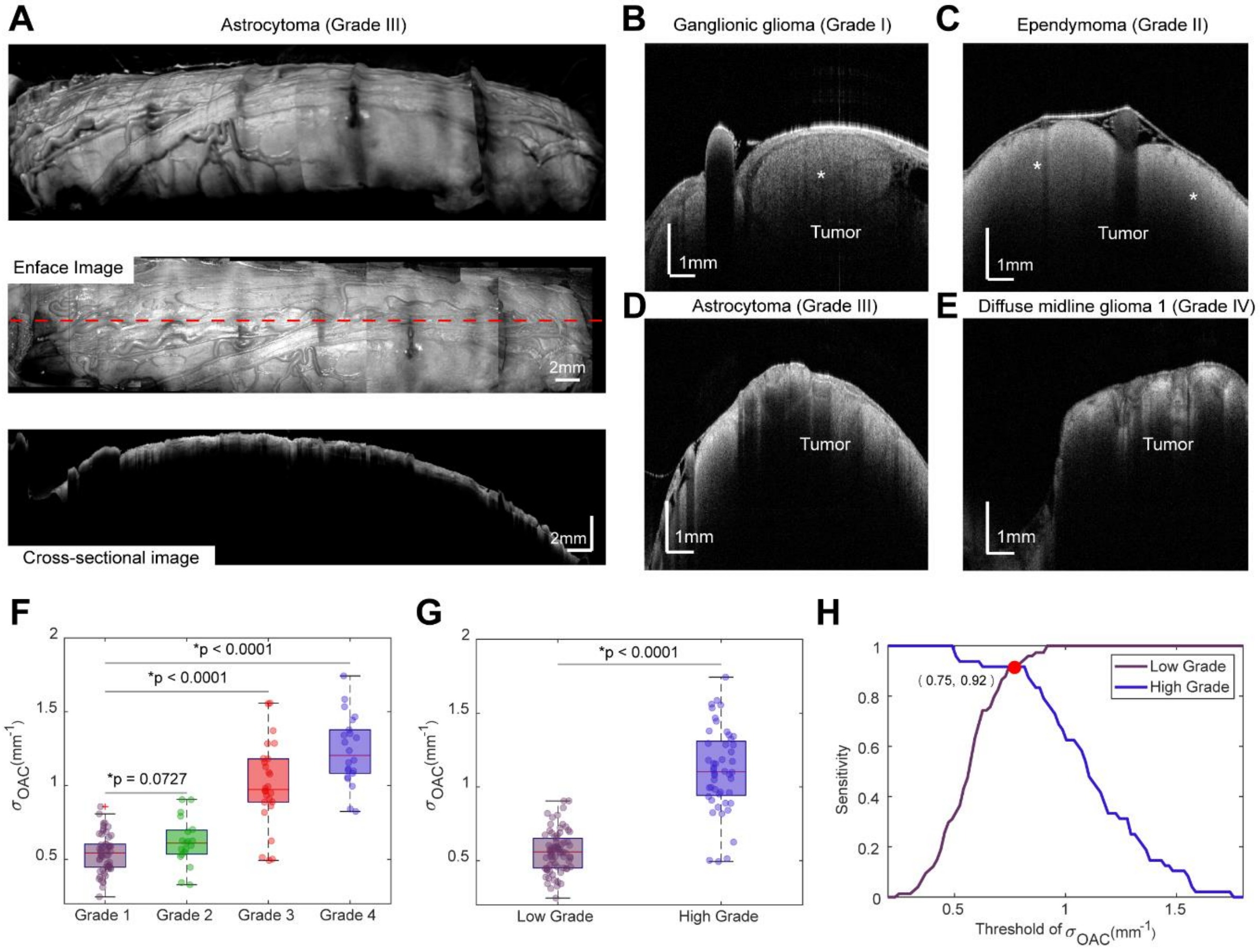

**Fig.5. FACT-ROCT Structural Images and Standard Deviation Analysis of Optical Attenuation Coefficients (OAC) for Different Grades of Spinal Cord Gliomas:** (A) 3D OCT structural image, enface image and cross-sectional image of an Astrocytoma (Grade III), the red dashed line marks the position of cross-sectional image. (B) - (E) the typical OCT B-scan images of Spinal Cord Gliomas with Different Grades. (F)Box plot showing the distribution of the standard variance of OAC for different grades of gliomas, showing that the standard deviation of OAC increases with higher grades. (G) Box plot comparing the standard variance of OAC between low-grade (Grade I and II) and high-grade (Grade III and IV) gliomas. (H) Sensitivity analysis of standard variance of OAC threshold values for distinguishing between low-grade and high-grade gliomas, achieving over 90% accuracy at the threshold of 0.75 mm$^{-1}$. For all box plots, the center lines represent the median values. The boxes extend from the lower quartile to the upper quartile, and the whiskers represent 1.5 times the interquartile range.

## 2.6 OCT angiography images for Spinal Cord Gliomas with Different Grades

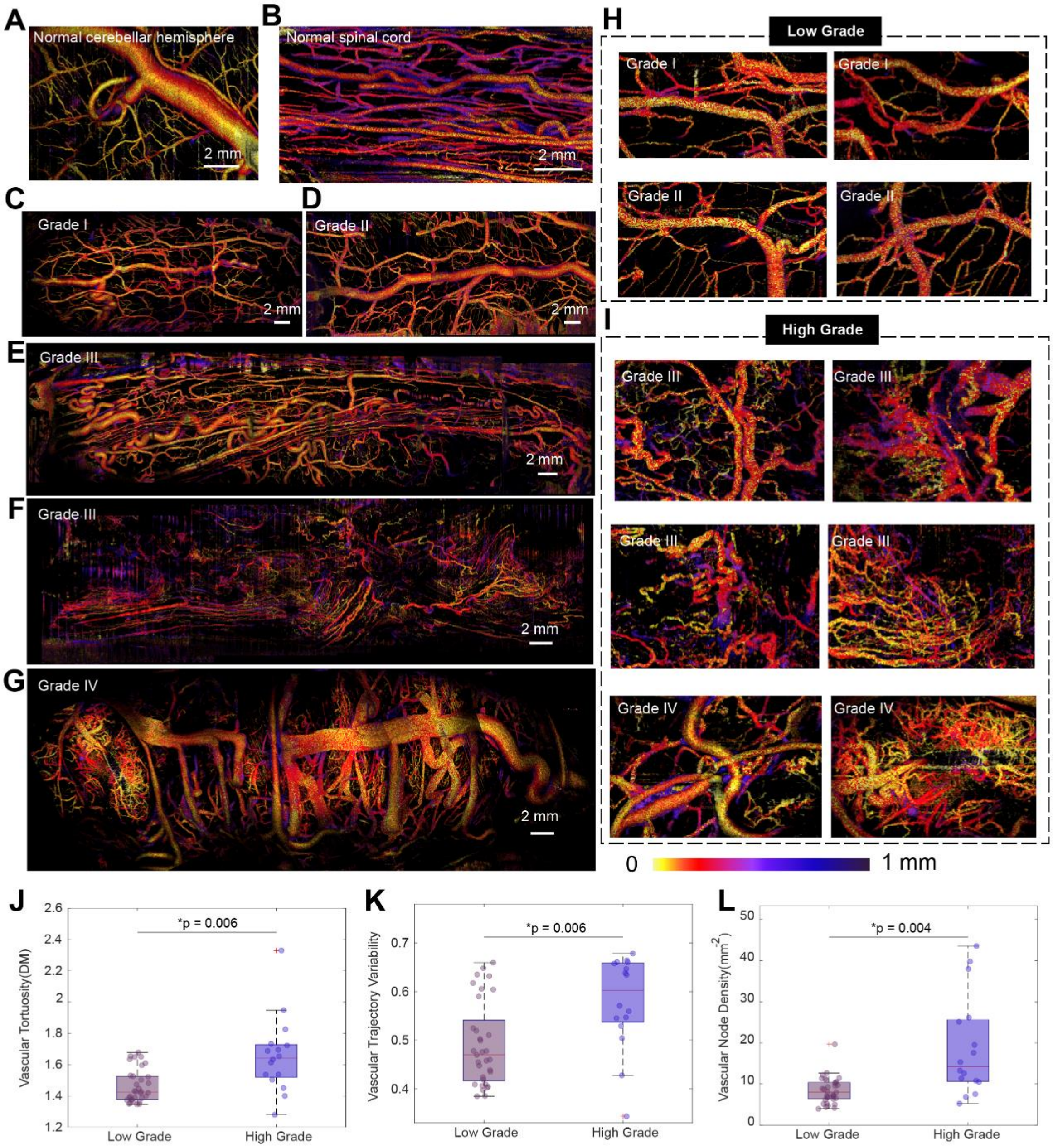

**Fig. 6 FACT-ROCT angiography results for the normal tissue and Spinal Cord Gliomas with Different Grades：** (A) The OCTA image of a normal cerebellar hemisphere reveals prominent draining vessels alongside smaller vessels. (B) The OCTA image of the normal thoracic spinal cord displays a structured vascular network with vessels oriented along its long axis, highlighting the organized nature of healthy spinal cord vasculature (C, D) OCTA images of Grade I and Grade II spinal cord gliomas, respectively. illustrate relatively regular and moderately dense surface vasculature, characteristic of low-grade gliomas. (E, F) The OCTA image of a WHO Grade III spinal cord glioma highlights an increased density and heterogeneity of the surface vasculature compared to lower-grade tumors, suggesting higher angiogenic activity. (G) The OCTA image of a WHO Grade IV spinal cord glioma shows a markedly dense and heterogeneous vascular network, with extensive microvascular proliferation and a mix of large draining vessels and irregular small vessels, indicative of high-grade gliomas. (H) Enlarged OCTA images of low-grade gliomas (Grade I and II) reveal moderately dense, regular vasculature with low heterogeneity, reinforcing the less aggressive nature of these tumors. (I) Enlarged OCTA images of high-grade gliomas (Grade III and IV) display dense, irregular vasculature with significant heterogeneity and pronounced microvascular proliferation, reflecting the aggressive angiogenic profile of high-grade tumors. (J)-(L) Quantitative analysis of vascular complexity in high- and low-grade spinal cord gliomas, including vascular tortuosity, vascular trajectory,

variability, and vascular node density, with high-grade gliomas showing higher values with significant differences ($p < 0.006$)

We also utilized FACT-ROCT angiography to investigate the vascular characteristics of normal tissues and spinal cord gliomas of various grades. The OCTA images provide insightful comparisons between healthy and tumorous tissues, emphasizing the variations in vascular structures associated with different tumor grades. In normal tissues, the cerebellar hemisphere (Fig. 6A) exhibits prominent draining vessels alongside smaller vessels, while the thoracic spinal cord (Fig. 6B) shows a structured vascular network oriented along its long axis, indicating organized vasculature. Low-grade gliomas, represented by Grade I (Fig. 6C) and Grade II (Fig. 6D), display relatively regular and moderately dense surface vasculature, characteristic of their less aggressive nature. In contrast, the OCTA images of high-grade gliomas, such as Grade III (Fig. 6E, F) and Grade IV (Fig. 6G), reveal significantly increased density and heterogeneity in the vascular network. Grade III gliomas show a higher density and varied surface vasculature, indicating elevated angiogenic activity, whereas Grade IV gliomas demonstrate a markedly dense and heterogeneous network, with extensive microvascular proliferation and a mix of large draining vessels and irregular small vessels, reflecting the aggressive nature of these tumors. Enlarged images of low-grade gliomas (Fig. 6H) further emphasize the moderately dense, regular vasculature with low heterogeneity, reinforcing their less aggressive behavior. Conversely, high-grade gliomas (Fig. 6I) present dense, irregular vasculature with significant heterogeneity and pronounced microvascular proliferation, highlighting their aggressive angiogenic profile. Quantitative analyses of vascular complexity parameters further support these visual observations (Methods for more details). The vascular tortuosity (Fig. 6J), vascular trajectory variability (Fig. 6K), and vascular node density (Fig. 6L) metrics all indicate significantly higher complexity in high-grade gliomas compared to low-grade gliomas, with statistically significant differences ($p = 0.006$ for tortuosity and trajectory variability, and $p = 0.004$ for node density). This quantitative evidence underscores the potential of FACT-ROCT angiography in distinguishing between low-grade and high-grade spinal cord gliomas based on their vascular characteristics.

## 3. Discussion

In this study, we present a first-in-human study of the multifunctional Optical Coherence Tomography technology, FACT-ROCT, for intraoperative imaging of spinal cord tumors. Spinal cord tumors pose a substantial clinical challenge due to their location and the critical functions of the spinal cord. Traditional intraoperative imaging modalities such as MRI, CT, and ultrasound, although valuable, often lack the necessary real-time, high-resolution capabilities required for optimal intraoperative guidance. FACT-ROCT has demonstrated artifact-free, high-resolution imaging of various types of spinal cord tumors, capturing micrometer structural and vascular information essential for surgical decision-making. The system has successfully imaged different grades of spinal cord tumors, including Grade I through Grade IV gliomas, providing clear differentiation between tumor grades based on structural and vascular characteristics.

Our FACT-ROCT system incorporates several key innovations pivotal for its success in a clinical setting. Firstly, the rapid adaptive focal tracking method, achieved through the integration of an ETL, ensures that the imaging focus is consistently maintained despite tissue movements. This real-time focus adjustment, with a response time of ~10 milliseconds, allows for high SNR and high-resolution imaging across varying tissue surfaces. Unlike traditional ETL-based OCT techniques that require multi-volume scanning or pre-scanning [29-31], our method involves a feedback loop for continuous focus refinement without needing prior knowledge of the object's shape. Therefore, FACT-ROCT is also able to addresses the lateral resolution degradation caused by inevitable physiological movements such as heartbeats during surgery. Although only the raster scanning protocol is used to demonstrated the effectiveness of our focus tracking method, more intricate scanning protocol [34] combined with the ETL can be employed in different situations, compensating for the limitation of ETL's response time. For instance, we can decompose a region into several long strip scans to achieve dynamic focusing along the fast axis. Additionally, our method is not restricted to any specific focal change technology, meaning that other high speed focal changing techniques can be applied to further enhance the dynamic focus tracking speed of FACT-ROCT [35, 36].

Secondly, the integration of a force-controlled 7-DOF robotic arm also enables continues and large-area automatic imaging, covering the entire spinal cord tumor region efficiently and safely. The robotic arm's capabilities, such as zero-force dragging and virtual wall control, enhance safety during intraoperative imaging by preventing unintentional probe movements and collisions. By automating the movement of the OCT probe based on previously obtained volumes, the system minimizes the need for manual adjustments and reduces the risk of human error. The imaging procedure is safe and well-tolerated, even in cases involving multiple consecutive imaging runs. No additional medications are required, and no imaging-related side effects are observed in the participating patients. In addition, the implementation of motion compensation techniques based on fast cross-sectional scanning addresses the deformation of OCT volumes caused by the physiological movements, maintaining image integrity and reliability during surgery.

In this first-in-human study, the FACT-ROCT system is used for intraoperative imaging on 21 patients with various spinal cord tumors, including spinal gliomas (grades I-IV), teratomas, hemangiomas, and lipomas, showing promising clinical outcomes. These imaging results demonstrate that FACT-ROCT system can provide micrometer-scale in vivo images of normal tissues, nerve boot and vessels, which are beneficial for the spinal cord tumor surgery. Additionally, the OCT structural images from different spinal cord tumors show that FACT-ROCT has the potential in differentiate the types of spinal cord tumors. Also, we also find high-grade gliomas exhibited more uneven scattering, correlating with increased heterogeneity and malignancy, while low-grade tumors showed more even scattering pattern.

Further quantitative analysis of OAC is conducted on the different grade spinal cord gliomas. It is worth to mention that our proposed adaptive focus tracking method ensures that the focus always remains at the same position inside the tissue, which guarantees the robustness of the OAC, even in the presence of tissue movement intraoperatively. The OAC results for *in situ* spinal cord gliomas are significantly different from those reported for ex vivo brain gliomas in previous research [22], possibly due to factors such as the continuous blood supply in living tissues. We find it difficult to directly distinguish between different grades of tumors and normal tissue just based on the OAC, while the standard deviation of the attenuation coefficient increases with higher-grade gliomas. Using the standard deviation of OAC as a physical marker, the system achieved over 90% accuracy in distinguishing high- from low-grade gliomas intraoperatively at a threshold of 0.75 $mm^{-1}$. This high diagnostic accuracy indicates FACT-ROCT's potential in providing critical real-time information for surgical decision-making.

In addition to structural imaging, FACT-ROCT also provides detailed vascular imaging. Unlike previous clinical studies on brain tumor vascular imaging [28], which only visualized blood vessels in B-scan images. FACT-ROCT, utilizing the robotic arm and the mentioned innovations, enables vascular imaging across the entire tumor area. It can cover a region of 70 mm * 13 mm * 10 mm within 2 minutes. This capability enables the creation of comprehensive vascular maps, essential for planning safe resections by avoiding critical blood vessels and minimizing intraoperative bleeding. Besides, our findings reveal that higher-grade tumors exhibit greater vascular tortuosity, a marker of their aggressive nature. These observations offer valuable information for surgical planning and may also contribute to a better understanding of tumor biology.

Despite the promising results, the FACT-ROCT system also has some limitations. Firstly, the imaging depth of OCT is limited to 1-2 mm, which may be insufficient for larger or deeply situated tumors. However, when FACT-ROCT seamlessly integrates into existing surgical workflows, the depth limitation becomes less of a concern, as the depth of surgical excision typically aligns with the 1-2 mm range. Additionally, the proposed motion compensation method is only suited for object movement along the longitudinal axis, which is also the most common scenario during surgery, as the robotic arm can ensure that the light is approximately perpendicular to the object surface. Moreover, by employing faster MHz swept-source lasers [37, 38], the imaging speed of the FACT-ROCT system can be further improved, but higher imaging speeds usually also mean lower sensitivity, lower imaging depth, and higher costs. Future designs will consider using the 10.3 MHz swept-source laser that we previously developed [39]. Additionally, this study involved a small sample size of patients with spinal cord tumors. More studies are necessary to validate these findings and assess the generalizability of FACT-ROCT across different tumor types and patient populations. Longitudinal studies can also evaluate the impact of FACT-ROCT-guided surgeries on long-term patient outcomes, such as recurrence rates and functional recovery.

In conclusion, the introduction of the FACT-ROCT system represents a significant advancement in the intraoperative imaging of spinal cord tumors. By addressing critical challenges in focus tracking, motion compensation, and large-scale scanning, our work enhances the precision and applicability of OCT in a surgical setting. The ability to perform high-resolution structural and vascular imaging in real-time improves the accuracy of tumor grading and informs surgical strategies, potentially enhancing patient outcomes. Future advancements and larger studies will further elucidate its role in clinical practice and solidify its position as a vital tool in neurosurgical oncology. FACT-ROCT also holds the potential to enhance the precision of artificial spinal cord implant surgeries and expand its applications to various tumor surgeries across different departments.

## 4. Materials and methods

### 4.1 Study design

The objective of this study was to evaluate the application of a multifunctional OCT system, including both structural and vascular imaging, in imaging spinal cord tumors in patients undergoing surgical resection. Acknowledging the limitations of existing imaging technologies in providing high-resolution, real-time images during spinal cord tumor surgeries, we developed the FACT-ROCT system. No additional medications were necessary for FACT-ROCT acquisitions. To ensure the safety of the entire experiment, we used a non-contact imaging approach with a high-resolution, variable-focus OCT probe mounted on a robotic arm, incorporating fast focus tracking and motion compensation methods to achieve artifact-free imaging during surgery. The primary objective was to evaluate the system's effectiveness in providing micrometer-resolution in vivo structural images of various spinal cord tumors and vascular imaging results, as well as its potential to enhance diagnostic accuracy and guide surgical decision-making. The study enrolled 21 patients with spinal cord tumors, including 13 with spinal gliomas (grades I to IV) and 8 with other tumors or lesions such as teratomas. The selected tumors were all indicated by MRI to protrude from the surface of the spinal cord, and their tumor types were subsequently determined through H&E pathological examination. The OCT structural and vascular images, along with the quantitative imaging findings were presented. The OAC and vascular tortuosity of different grade spinal cord gliomas were calculated to demonstrate the increased heterogeneity with higher grade tumors, and to indicate that FACT-ROCT has the capability to serve as an intraoperative real-time imaging method for tumor grading. Participants were enrolled through a clinical study protocol approved by the Tsinghua Changgung Hospital Ethics Committee (protocol #23703-4-01). No randomization and blinding were performed.

## 4.2 FACT-ROCT system

The FACT-ROCT is based on a home-built robotic SS-OCT system, as shown in Fig. S1. A 200 kHz swept-source laser (Axsun Technology) with a center wavelength of 1310 nm and a sweeping bandwidth of 100 nm was incorporated into the system. A 1:99 fiber coupler was used, where 1% of the light passed through a circulator into a Fiber Bragg Grating (FBG, $\lambda_0$=1327 nm, reflectivity = 99%, $\Delta\lambda$=0.4 nm, Guangxin Technology) to achieve stable A-line wavelength triggering, ensuring the phase stability of the system. The remaining 99% of the light was further split by a 10:90 coupler, with 10% going to the reference arm and 90% to the sample arm. The light was then recombined using a 50:50 coupler and directed into a 1.6 GHz balanced detector (ABD-1.6G-A, Wuhan Optolabs). In the sample arm, the light first passed through an ETL (EL-10-30-C, Optotune), and then a 2D galvanometer (S8107, Sunny Technology) was used to achieve two-dimensional lateral scanning. After scanning, the light passed through a scan lens (LSM04, Thorlabs) with a focal length of 54 mm, focusing on the spinal cord tumor tissue. A 632 nm red light was introduced into the sample arm to guide the scanning. The generated A-line trigger was sent to a digital-to-analog (DA) card (USB 3020, ART Technology) as a clock signal, and frame triggers and galvanometer control analog signals were generated by this DA card. The interference signal was acquired by a 12-bit acquisition card (ATS9373, AlazarTech) and then sent to a GPU for post-processing. The real-time OCT B-scan signal drove a PCIe DA card (PCIe 6353, National Instruments) to generate voltage signals for ETL focal changing. The lateral scan range of the system for a single subject was 13 mm × 13 mm, and the axial scan range was 5 mm. The lateral and axial resolutions of the system were ~25 μm and ~10 μm in tissue, respectively. The time to achieve structural imaging for a single subject was 2.5 seconds, and vascular imaging took 10 seconds. The system achieved a sensitivity of 105 dB with a sample power of 10 mW and the sensitivity roll of is -0.2dB per mm.

The entire sample arm was mounted on a seven-axis force-controlled medical robotic arm (ER3 Pro-Med). The robotic arm has a collision detection sensitivity of less than 3N, with a real-time refresh rate of 1 kHz. It also supports near zero-force lightweight dragging, making it easy for medical staff to operate. Additionally, the robotic arm is equipped with a real-time controllable virtual wall function, which can restrict the movement range of the tool. After mounting the OCT sample arm onto the robotic arm using custom mechanical parts, a two-step calibration method was applied to match the OCT coordinate system with the robotic arm's coordinate system [40-42], allowing the robotic arm to move guided by the OCT images. To compensate for imaging distortion caused by heartbeat during surgery, a cross-sectional scan is performed before volumetric imaging to obtain the surface contour of the object. This contour is used to compensate for object motion during volumetric imaging. Furthermore, after acquiring the first compensated volumetric image, the robotic arm plans the next scan position based on the last 200 frames of data from the previous volumetric scan, repeating this process to scan all areas of the spinal cord tumor (SFig.5).

During OCT volumetric imaging, an adaptive focal tracking strategy based on real-time OCT B-scan images is used to compensate for the defocus caused by object motion and curvature. Specifically, after obtaining the first OCT B-scan image, the surface positions of the 10 central A-lines of the B-scan are identified and averaged using a rapid surface detection algorithm [43]. Based on the continuity of the object, the ETL is then used to adjust the focal point of the next B-scan to ~400 μm below the surface of the object. This method has a response time of 10 ms. By using this approach, where the previous B-scan drives the precise focusing of the next B-scan, almost all B-scans of the object can be maintained in the high-resolution region.

## 4.3 Optical attenuation coefficient and its Standard Deviation

The quantitative analysis of OCT structural images for different grades of spinal cord gliomas is based on their attenuation coefficients. Due to light attenuation in tissue, and considering the effects of defocusing and system fall-off, the average OCT signal intensity can be expressed as: [44]:

$$< A(z) >= A_0 \cdot t(z - z_f) \cdot h(z) \cdot e^{(-\mu \cdot z)} + noise \qquad (1)$$

Where $A_0$ is a constant, $z$ is the depth pixel in OCT images, $t(z - z_f)$ represents the defocus-related term. As for the adaptive focal tracking strategy, the focal point is always located 400 μm beneath the surface of the object, which is fixed to avoid inconsistencies in attenuation coefficients caused by focal changes in traditional methods. And $h(z)$ is the sensitivity roll-off factor term. When calculating the attenuation coefficient, regions with obvious surface blood vessel occlusion and non-spinal cord areas at the edges are first excluded. For the remaining valid A-lines, the surface contour of the OCT image is obtained using an automatic detection algorithm. Then, the signals below the surface from five consecutive A-lines in the lateral direction are averaged to obtain the average light intensity <A(z)>. From this dataset, several segments at equal distances below the surface are randomly selected for fitting using Eq. (1) to calculate the attenuation coefficient $\mu$. For one B-scan, 10 A-lines are randomly selected, resulting in 10 values of the attenuation coefficient. Then, the standard deviation of these attenuation coefficients $\sigma_\mu$ is calculated as:

$$\sigma_\mu = \sqrt{\frac{\sum_{i=1}^{10} (\mu_i - \bar{\mu})^2}{9}} \qquad (2)$$

For each case, approximately 1000 to 5000 B-scans were collected, depending on the tumor size, representing OCT image samples from different locations of the tumor. Some B-scans were randomly selected from this dataset, and the standard deviation of the corresponding attenuation coefficients was calculated. This standard deviation serves as a statistical parameter reflecting the heterogeneity of the attenuation coefficient.

## 4.4 Spinal cord tumor vascular imaging and vessel parameter analysis

Label-free vascular imaging of spinal cord tumors can be achieved by performing four consecutive B-scans at the same location, followed by applying a vascular algorithm based on correlation-coded eigen decomposition [40, 45, 46]. To remove vertical stripe artifacts caused by intraoperative motion, a Fourier domain spatial filtering method was employed. Details of the specific vascular processing algorithm can be found in the Supplemental Materials (SFig.6). Once the vascular imaging results are obtained, different depths of the vasculature are displayed using color mapping. To analyze the quantitative vascular parameters, the vascular results are projected into 2D enface images, and Vascular Tortuosity [47], Vascular Node Density, and Vascular Trajectory Variability are selected as statistical parameters to compare the differences between high- and low-grade spinal cord glioma vasculature. The average curvature of vessels in the region (also referred to as distance metric, DM) is defined as the total length of all vessels divided by the sum of the distances between the start and end points of all vessels. The density of vascular nodes is defined as the number of vascular branches per unit area. Vascular trajectory variability is defined as the standard deviation of the azimuth angles of the local directions of all vessels. During processing, the 2D projection of the vessels obtained from OCTA is first subjected to Gaussian filtering, Frangi filtering, and binarization. The binarized vessel map is then processed using morphological methods for denoising and subsequent extraction of the vessel skeleton, after which the vascular parameters are calculated. Firstly, the number of vascular branches, as shown in (SFig. 7A), is counted from the vascular skeleton diagram and divided by the area of the region to obtain the vascular node density, defined as:

$$VND = \frac{Number\ of\ nodes}{Area\ of\ region} \tag{3}$$

Next, all node positions of the vascular diagram are disconnected to generate a subgraph composed of a series of non-intersecting vessels. For each vessel in this subgraph, the total length $L_i$ and the distance between its start and end points $s_i = |\overrightarrow{AB}|$ are calculated, as illustrated in (SFig. 7B). Finally, the total lengths $L_i$ and distances $s_i$ of all vessels are summed, and the average vascular curvature is computed as:

$$DM = \frac{\sum L_i}{\sum s_i} \tag{4}$$

In the subgraph composed of non-intersecting vessels, the azimuth angles $\phi_i$ of each vessel at specified intervals are recorded, as shown in (SFig. 7C). The number of azimuth angles $\phi_i$ for all vessels in the region is tallied, with the most frequent $\phi_i$ identified as the primary flow direction $\phi_m$ of the vessels in that area. Then, for all azimuth angles that form an obtuse angle with this primary direction, they are normalized to $(\phi_m - \frac{\pi}{2},\ \phi_m + \frac{\pi}{2}]$ by $\phi_i' = \phi_i \pm \pi$. Subsequently, the standard deviation is calculated as the variability of vascular trajectories, expressed as:

$$VTV = \sqrt{\frac{\sum(\phi_i' - \overline{\phi_i'})^2}{n-1}} \tag{5}$$

where $n$ is the total number of azimuth angles involved in the statistics, and $\overline{\phi_i'}$ represents the mean of all normalized $\phi_i'$. From these parameters that reflect the complexity of the vessels, high-grade tumors with complex and disordered vascular structures can be distinguished from low-grade tumors with simpler and more organized vascular patterns.

## 4.5 Statistical analysis

In this study, a significance test was performed to evaluate the difference in the standard deviation of the attenuation coefficient ($\sigma_\mu$) between low-grade and high-grade spinal cord gliomas. A t-test was used to assess whether the difference in $\sigma_\mu$ between the two groups was statistically significant. Representative data from 13 patients with WHO Grade I, II, III, and IV gliomas were selected for analysis. From the thousands of B-scans obtained from each patient, approximately 20 B-scans containing tumor regions (around 1% of the total B-scans) were randomly selected for the calculation of $\sigma_\mu$. This resulted in 20 $\sigma_\mu$ values per patient. The low-grade and high-grade groups each contained approximately 120 $\sigma_\mu$ values. The sample size satisfied the assumptions for conducting a t-test, providing a statistical power of greater than 0.8 with a significance level of $\alpha = 0.01$. The t-test results led to the rejection of the null hypothesis (H0: the mean $\sigma_\mu$ for low-grade and high-grade tumors is equal) at a significance level of $\alpha = 0.01$, indicating a statistically significant difference in the standard deviation of the attenuation coefficient ($\sigma_\mu$) between the low-grade and high-grade tumors. To assess the differences in the three vascular parameters between the low-grade and high-grade groups, a t-test was also conducted. Each sample was obtained from a randomly selected subregion of the tumor area. On average, each case was divided into five subregions, and the parameters were calculated separately for each. Approximately 25 subregions were analyzed for both the low-grade and high-grade groups. The sample size met the requirements for performing a t-test with a significance level of $\alpha = 0.05$ and a statistical power of 0.8. The t-test results yielded p-values of 0.006, 0.006, and 0.004 for the three vascular parameters, respectively, confirming statistically significant differences between the low-grade and high-grade groups for all three parameters.

## 5. Data availability

All data supporting the findings of this study are available within this paper and its Supplementary information file. The raw OCT data, due to the clinical informed consent, are not available now. But the phantom data which used to validate the system is available from the corresponding authors.

## 6. Code availability

All code used in this study is available from the corresponding authors upon reasonable request.

## 7. References

1. E. Diaz, H. Morales, Spinal Cord Anatomy and Clinical Syndromes. *Seminars in Ultrasound, CT and MRI* **37**, 360–371 (2016).
2. C. S. Ahuja, J. R. Wilson, S. Nori, M. R. N. Kotter, C. Druschel, A. Curt, M. G. Fehlings, Traumatic spinal cord injury. *Nat Rev Dis Primers* **3**, 17018 (2017)
3. N. Sanai, M. S. Berger, Glioma Extent of Resection and Its Impact On Patient Outcome. *Neurosurgery* **62**, 753 (2008).
4. D. Kuhnt, A. Becker, O. Ganslandt, M. Bauer, M. Buchfelder, C. Nimsky, Correlation of the extent of tumor volume resection and patient survival in surgery of glioblastoma multiforme with high-field intraoperative MRI guidance. *Neuro-Oncology* **13**, 1339–1348 (2011).
5. S. Zausinger, B. Scheder, E. Uhl, T. Heigl, D. Morhard, J.-C. Tonn, Intraoperative Computed Tomography With Integrated Navigation System in Spinal Stabilizations. *Spine* **34**, 2919-2926 (2009).
6. O. M. Rygh, T. Selbekk, S. H. Torp, S. Lydersen, T. A. N. Hernes, G. Unsgaard, Comparison of navigated 3D ultrasound findings with histopathology in subsequent phases of glioblastoma resection. *Acta Neurochir (Wien)* **150**, 1033–1042 (2008).
7. M. H. A. Noureldine, N. Shimony, G. I. Jallo, “Malignant Spinal Tumors” in *Human Brain and Spinal Cord Tumors: From Bench to Bedside. Volume 2,* (Springer International Publishing, Cham, 2023) 1405, pp. 565–581.
8. V. Espina, Ed., "Tumor Staging and Grading: A Primer" in *Molecular Profiling: Methods and Protocols* (Humana Press, 2017), pp. 1–16.
9. H. Jaafar, Intra-Operative Frozen Section Consultation: Concepts, Applications and Limitations. *Malays J Med Sci.* **13**, 4-12 (2006).

10. M. L. Urken, J. Yun, M. P. Saturno, L. A. Greenberg, R. L. Chai, K. Sharif, M. Brandwein-Weber, Frozen Section Analysis in Head and Neck Surgical Pathology: A Narrative Review of the Past, Present, and Future of Intraoperative Pathologic Consultation. *Oral Oncology* **143**, 106445 (2023).

11. J. A. Nagy, H. F. Dvorak, Heterogeneity of the tumor vasculature: the need for new tumor blood vessel type-specific targets. *Clin Exp Metastasis*. **29**, 657–662 (2012).

12. D. N. Louis, A. Perry, P. Wesseling, D. J. Brat, I. A. Cree, D. Figarella-Branger, C. Hawkins, H. K. Ng, S. M. Pfister, G. Reifenberger, R. Soffietti, The 2021 WHO Classification of Tumors of the Central Nervous System: a summary. *Neuro Oncol.* **23**, 1231-1251 (2021).

13. J. W. Kernohan, R. F. Mabon, H. J. Svien, A. W. Adson. A simplified classification of the gliomas. *Proc Staff Meet Mayo Clin* **24**, 71-75 (1949).

14. C. Daumas-Duport, B. Scheithauer, J. O'Fallon, P. Kelly, Grading of astrocytomas: A simple and reproducible method. *Cancer* **62**, 2152–2165 (1988).

15. C. W. Primrose, E. M. Hecht, G. Roditi, C. J. François, J. H. Maki, C. L. Dumoulin, J. K. DeMarco, P. Douglas, MR Angiography Series: Fundamentals of Contrast-enhanced MR Angiography. *RadioGraphics* **41**, E138–E139 (2021).

16. H. G. J. Kortman, E. J. Smit, M. T. H. Oei, R. Manniesing, M. Prokop, F. J. A. Meijer, 4D-CTA in Neurovascular Disease: A Review. *AJNR Am J Neuroradiol* **36**, 1026–1033 (2015).

17. T. J. Bennett, D. A. Quillen, R. Coronica, Fundamentals of Fluorescein Angiography. *Insight* **41**, 5-11 (2016).

18. B. E. Bouma, J. F. De Boer, D. Huang, I.-K. Jang, T. Yonetsu, C. L. Leggett, R. Leitgeb, D. D. Sampson, M. Suter, B. J. Vakoc, M. Villiger, M. Wojtkowski, Optical coherence tomography. *Nat Rev Methods Primers* **2**, 79 (2022).

19. W. Drexler, U. Morgner, R. K. Ghanta, F. X. Kärtner, J. S. Schuman, J. G. Fujimoto, Ultrahigh-resolution ophthalmic optical coherence tomography. *Nat Med* **7**, 502–507 (2001)..

20. B. E. Bouma, M. Villiger, K. Otsuka, W. Oh, Intravascular optical coherence tomography. *Biomed. Opt. Express* **8**, 2660-2686 (2017).

21. K. Yashin, M. M. Bonsanto, K. Achkasova, A. Zolotova, A.-M. Wael, E. Kiseleva, A. Moiseev, I. Medyanik, L. Kravets, R. Huber, R. Brinkmann, N. Gladkova, OCT-Guided Surgery for Gliomas: Current Concept and Future Perspectives. *Diagnostics* **12**, 335 (2022).

22. K. Bizheva, W. Drexler, M. Preusser, A. Stingl, T. Le, H. Budka, A. Unterhuber, B. Hermann, B. Povazay, H. Sattmann, A. F. Fercher, Imaging ex vivo healthy and pathological human brain tissue with ultra-high-resolution optical coherence tomography. *J. Biomed. Opt.* **10**, 011006 (2005).

23. C. Kut, K. L. Chaichana, J. Xi, S. M. Raza, X. Ye, E. R. McVeigh, F. J. Rodriguez, A. Quiñones-Hinojosa, X. Li, Detection of human brain cancer infiltration ex vivo and in vivo using quantitative optical coherence tomography. *Sci. Transl. Med.* **7**, 292ra100-292ra100 (2015).

24. M. Almasian, L. S. Wilk, P. R. Bloemen, T. G. van Leeuwen, Pilot feasibility study of in vivo intraoperative quantitative optical coherence tomography of human brain tissue during glioma resection. *J. Biophotonics* **12**, e201900037 (2019).

25. P. Kuppler, P. Strenge, B. Lange, S. Spahr-Hess, W. Draxinger, C. Hagel, D. Theisen-Kunde, R. Brinkmann, R. Huber, V. Tronnier, M. M. Bonsanto, Microscope-integrated optical coherence tomography for in vivo human brain tumor detection with artificial intelligence. *Journal of Neurosurgery,* 1–9 (2024).

26. W. Draxinger, N. Detrez, P. Strenge, V. Danicke, D. Theisen-Kunde, L. Schützeck, S. Spahr-Hess, P. Kuppler, J. Kren, W. Wieser, M. Mario Bonsanto, R. Brinkmann, R. Huber, Microscope integrated MHz optical coherence tomography system for neurosurgery: development and clinical in-vivo imaging. *Biomed. Opt. Express* **15**, 5960 (2024).

27. B. J. Vakoc, R. M. Lanning, J. A. Tyrrell, T. P. Padera, L. A. Bartlett, T. Stylianopoulos, L. L. Munn, G. J. Tearney, D. Fukumura, R. K. Jain, Three-dimensional microscopy of the tumor microenvironment in vivo using optical frequency domain imaging. *Nat Med* **15**, 1219–1223 (2009).

28. H. Ramakonar, B. C. Quirk, R. W. Kirk, J. Li, A. Jacques, C. R. P. Lind, R. A. McLaughlin, Intraoperative detection of blood vessels with an imaging needle during neurosurgery in humans. *Sci. Adv.* **4**, eaav4992 (2018).

29. Y. Li, P. Tang, S. Song, A. Rakymzhan, R. K. Wang, Electrically tunable lens integrated with optical coherence tomography angiography for cerebral blood flow imaging in deep cortical layers in mice. *Opt. Lett.* **44**, 5037 (2019).

30. J. P. Su, Y. Li, M. Tang, L. Liu, A. D. Pechauer, D. Huang, G. Liu, Imaging the anterior eye with dynamic-focus swept-source optical coherence tomography. *J. Biomed. Opt* **20**, 126002 (2015).

31. J. Liu, Y. Shi, Z. Gong, Y. Zhang, R. K. Wang, Adaptive contour-tracking to aid wide-field swept-source optical coherence tomography imaging of large objects with uneven surface topology. *Biomed. Opt. Express* **15**, 4891 (2024).

32. J. Zhao, Y. Winetraub, L. Du, A. Van Vleck, K. Ichimura, C. Huang, S. Z. Aasi, K. Y. Sarin, A. De La Zerda, Flexible method for generating needle-shaped beams and its application in optical coherence tomography. *Optica* **9**, 859 (2022).

33. Z. Ding, H. Ren, Y. Zhao, J. S. Nelson, Z. Chen, High-resolution optical coherence tomography over a large depth range with an axicon lens. *Opt. Lett.* **27**, 243 (2002).

34. M. Draelos, C. Viehland, R. P. McNabb, A. N. Kuo, J. A. Izatt, Adaptive point-scan imaging beyond the frame rate–resolution limit with scene-reactive scan trajectories. *Optica* **9**, 1276 (2022).

35. S. Kang, M. Duocastella, C. B. Arnold, Variable optical elements for fast focus control. *Nat. Photonics* **14**, 533–542 (2020).

36. T. Chakraborty, B. Chen, S. Daetwyler, B.-J. Chang, O. Vanderpoorten, E. Sapoznik, C. F. Kaminski, T. P. J. Knowles, K. M. Dean, R. Fiolka, Converting lateral scanning into axial focusing to speed up three-dimensional microscopy. *Light Sci Appl* **9**, 165 (2020).

37. M. Siddiqui, A. S. Nam, S. Tozburun, N. Lippok, C. Blatter, B. J. Vakoc, High-speed optical coherence tomography by circular interferometric ranging. *Nat. Photon*. **12**, 111–116 (2018).

38. M. Göb, T. Pfeiffer, T. Pfeiffer, W. Draxinger, S. Lotz, J. P. Kolb, R. Huber, Continuous spectral zooming for in vivo live 4D-OCT with MHz A-scan rates and long coherence. *Biomed. Opt. Express* **13**, 713–727 (2022).

39. C. Wang, Z. Yin, B. He, Z. Chen, Z. Hu, Y. Shi, X. Zhang, N. Zhang, L. Jing, G. Wang, P. Xue, Polarization-isolated stretched-pulse mode-locked swept laser for 10.3-MHz A-line rate optical coherence tomography. *Opt. Lett.* **48**, 4025 (2023).

40. B. He, Y. Zhang, Z. Meng, Z. He, Z. Chen, Z. Yin, Z. Hu, Y. Shi, C. Wang, X. Zhang, N. Zhang, G. Wang, P. Xue, Whole Brain Micro-Vascular Imaging Using Robot Assisted Optical Coherence Tomography Angiography. *IEEE J. Select. Topics Quantum Electron.* **29**, 1–9 (2023).

41. B. He, Y. Zhang, L. Zhao, Z. Sun, X. Hu, Y. Kang, L. Wang, Z. Li, W. Huang, Z. Li, G. Xing, F. Hua, C. Wang, P. Xue, N. Zhang, Robotic-OCT guided inspection and microsurgery of monolithic storage devices. *Nat Commun* **14**, 5701 (2023).

42. Y. Huang, Y. Huang, Y. Huang, X. Li, X. Li, J. Liu, Z. Qiao, J. Chen, Q. Hao, Q. Hao, Robotic-arm-assisted flexible large field-of-view optical coherence tomography. *Biomed. Opt. Express* **12**, 4596–4609 (2021).

43. B. He, Y. Zhang, Z. Meng, Z. He, C. Wang, Z. Chen, Z. Yin, Z. Hu, Y. Shi, N. Zhang, W. Zhang, G. Wang, P. Xue, Optical coherence tomography angiography with adaptive multi-time interval. *J. Biophotonics*, doi: 10.1002/jbio.202200340 (2023).

44. P. Gong, M. Almasian, G. van Soest, D. M. de Bruin, T. G. van Leeuwen, D. D. Sampson, D. J. Faber, Parametric imaging of attenuation by optical coherence tomography: review of models, methods, and clinical translation. *J. Biomed. Opt.* **25**, 040901 (2020).

45. S. Yousefi, Z. Zhi, R. K. Wang, Eigendecomposition-based clutter filtering technique for optical microangiography. *IEEE transactions on biomedical engineering* **58**, 2316–2323 (2011).

46. J. Enfield, E. Jonathan, M. Leahy, In vivo imaging of the microcirculation of the volar forearm using correlation mapping optical coherence tomography (cmOCT). *Biomed. Opt. Express* **2**, 1184-1193 (2011).

47. E. Bullitt, G. Gerig, S. M. Pizer, W. Lin, S. R. Aylward, Measuring tortuosity of the intracerebral vasculature from MRA images. *IEEE Transactions on Medical Imaging* **22**, 1163–1171 (2003).

## 8. Acknowledgments

This research was sponsored by the National Natural Science Foundation of China 61975091, 62475131, 61905015; Tsinghua Precision Medicine Foundation; Beijing Natural Science Foundation L246072, 7244448; Beijing Nova Program (0220484179, Z191100001119039). We would like to thank Xinyi Tang from Tsinghua University for assistance with image formatting.

## 9. Author contributions

B.H.: conceptualizing and building the system, creating and maintaining the software, designing all experiments, writing the original article; Y.Z.Y: writing and editing the original article, designing clinical experiments; Y.J.S.: building the system, maintaining the software, editing the article; Z.M: designing clinical experiments; Z.C.Y and Z.Y.C: data visualization; Z.W.H and R.Z.X: maintaining the system; L.K.J, Y.L, Z.X.S, W.T.M, Y.T.W and D.L: Surgical and Sample preparation; N.Z: funding acquisition; G.H.W: manuscript reviewing, funding acquisition, and supervising the project; P.X.: manuscript reviewing and editing, funding acquisition, and supervising the project;

## 10. Competing interests

The authors declare no competing interests.

# Supplemental File: First-in-human spinal cord tumor imaging with fast adaptive focal tracking robotic-OCT

Bin He[1,2#], Yuzhe Ying[3#], Yejiong Shi[1,2#], Zhe Meng[3], Zichen Yin[1,2], Zhengyu Chen[1,2], Zhangwei Hu[1,2], Ruizhi Xue[1,2], Linkai Jing[3], Yang Lu[3], Zhenxing Sun[3], Weitao Man[3], Youtu Wu[3], Dan Lei[3], Ning Zhang[4], Guihuai Wang[3*] and Ping Xue[1,2*]

*[1]State Key Laboratory of Low-dimensional Quantum Physics and Department of Physics, Tsinghua University*

*[2]Frontier Science Center for Quantum Information, Beijing, China*

*[3]Department of Neurosurgery, Beijing Tsinghua Changgung Hospital, School of Clinical Medicine and Institute of Precision Medicine, Tsinghua University, Beijing, 102218, China.*

*[4]Institute of Forensic Science, Ministry of Public Security, Beijing, 100038, China*

*[#] These authors contributed equally.*

* *youngneurosurgeon@163.com*

* *xuep@tsinghua.edu.cn*

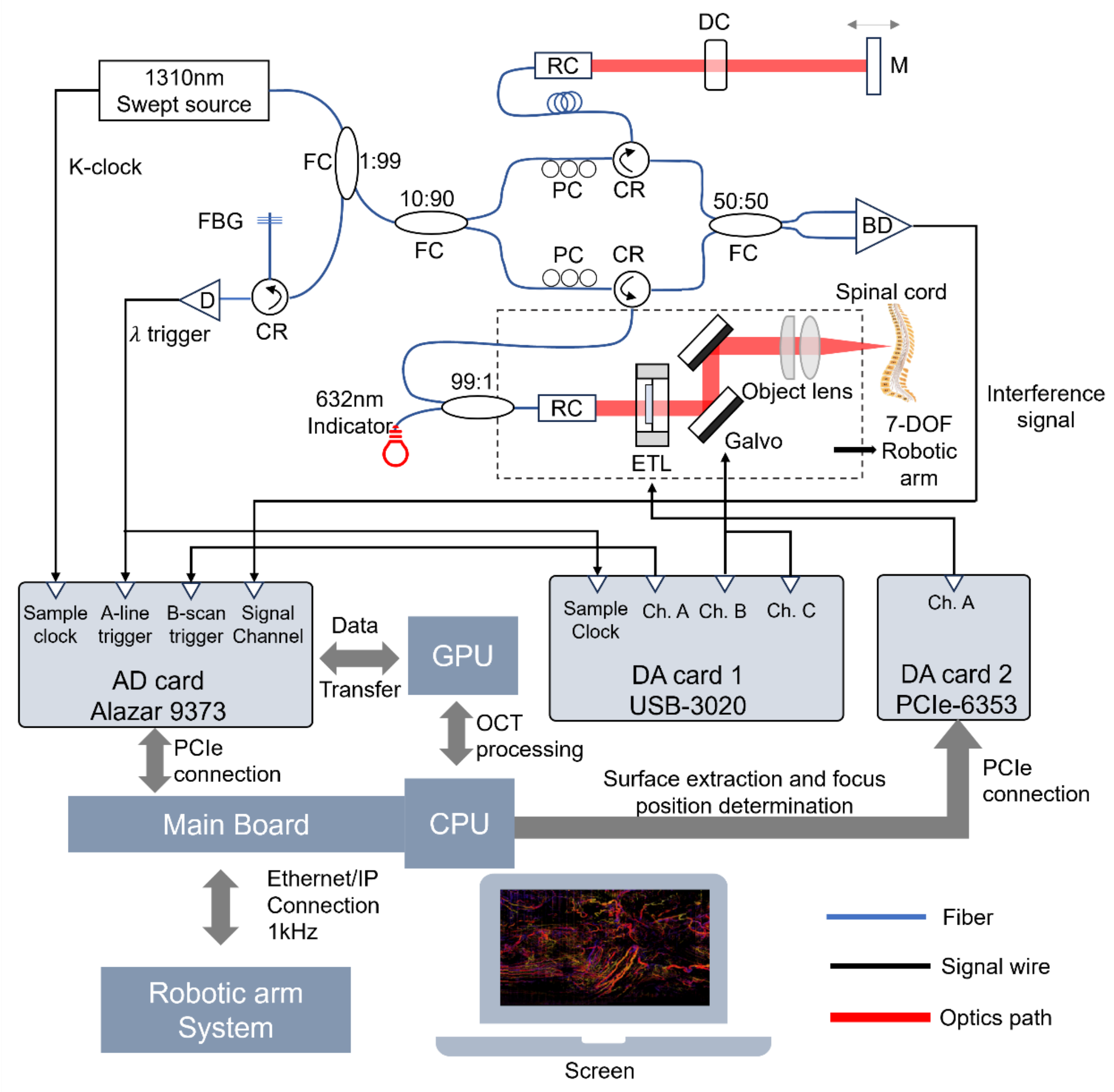

**Sfig.1: Schematic of FACT-ROCT system**. Abbreviations: FC, fiber coupler; FBG, fiber bragg grating; ETL: electrically tunable lens; RC, reflective collimator; PC, polarization controller; BD: balanced detector.

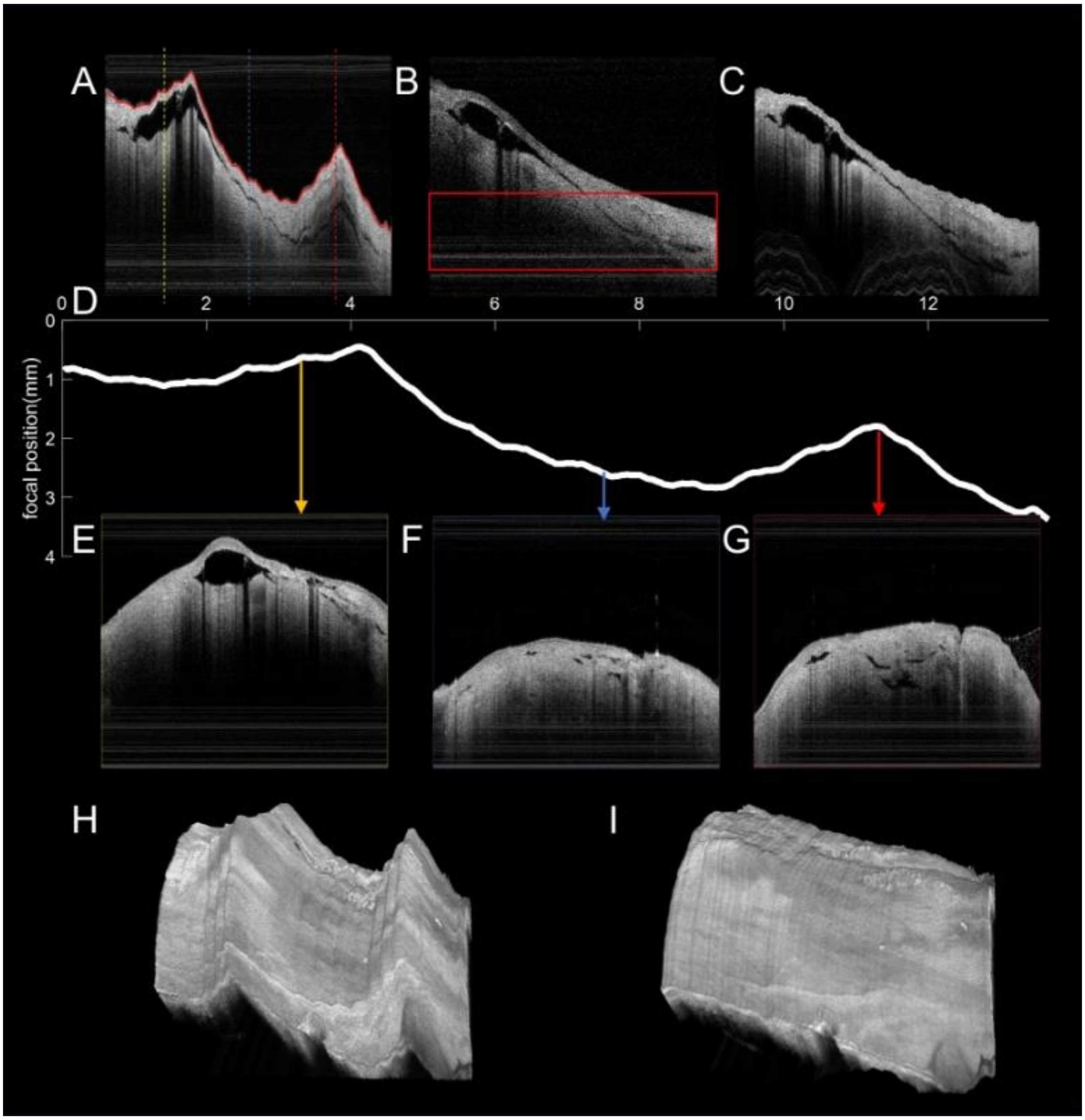

**SFig.2: Focus Tracking Demonstration of the FACT-ROCT System under Longitudinal Jitter of Up to 3 mm During Intraoperative Imaging:** (A) OCT B-scan along the slow axis scanning direction of the galvanometer, showing significant longitudinal jitter. (B) OCT B-scan in the slow-axis direction under fast cross-scanning, showing no motion artifacts, but only the region within the red box is within the depth of focus. (C) OCT B-scan along the slow-axis direction after motion compensation, showing the removal of artifacts and all regions being within the depth of focus. (D) Focal position correction curves over time used for adaptive focusing, indicating that the jitter of the object reached approximately 3 mm. (E, F, G) OCT B-scan images corresponding to the arrowed positions in panel D, showing that B-scan at different depths is within the depth of focus, and the signal-to-noise ratio is high in all cases. (H) OCT volume before motion compensation, and (I) OCT volume after motion compensation.

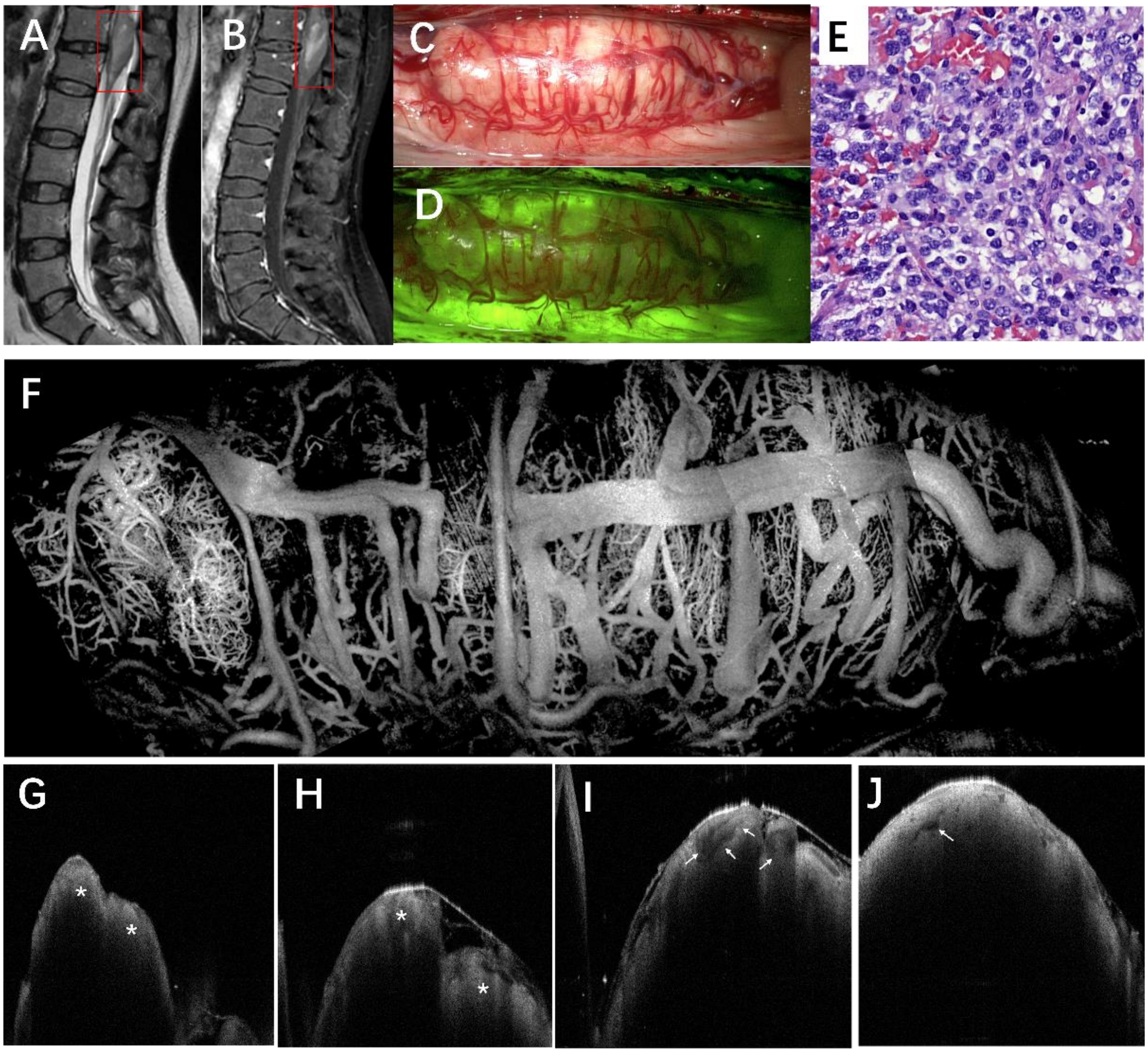

**SFig.3 A case of diffuse midline glioma:** (A-B) Preoperative MRI of the patient shows the tumor located at the T12-L1 vertebral level, occupying most of the spinal cord, causing fusiform swelling and surface breakthrough. Contrast-enhanced MRI shows significant and also heterogeneous signals. (C) Intraoperative macroscopic view of the tumor shows that the tumor is prominently protruding from the surface of the spinal cord, with a rich blood supply and a high density of surface vessels. (D) Intraoperative fluorescence imaging of the tumor after sodium fluorescein injection shows fluorescein enrichment on the left side of the tumor. (E) Postoperative pathology image of the tumor specimen shows densely distributed tumor cells of various shapes, with some cells displaying a plump spindle shape and marked atypia. Microvascular proliferation and hemorrhage are also observed, characteristic of WHO Grade IV glioma. (F) OCTA vascular imaging shows dense and irregular tumor vasculature, with large surface vessels corresponding to the macroscopic image in panel C. (G-J) Selected OCT B-scan structural images from the tumor’s rostral to caudal ends show heterogeneous optical signals within the tumor (indicated by "*"), with some vessels traversing the tumor surface or penetrating through the tumor (as indicated by white arrows).

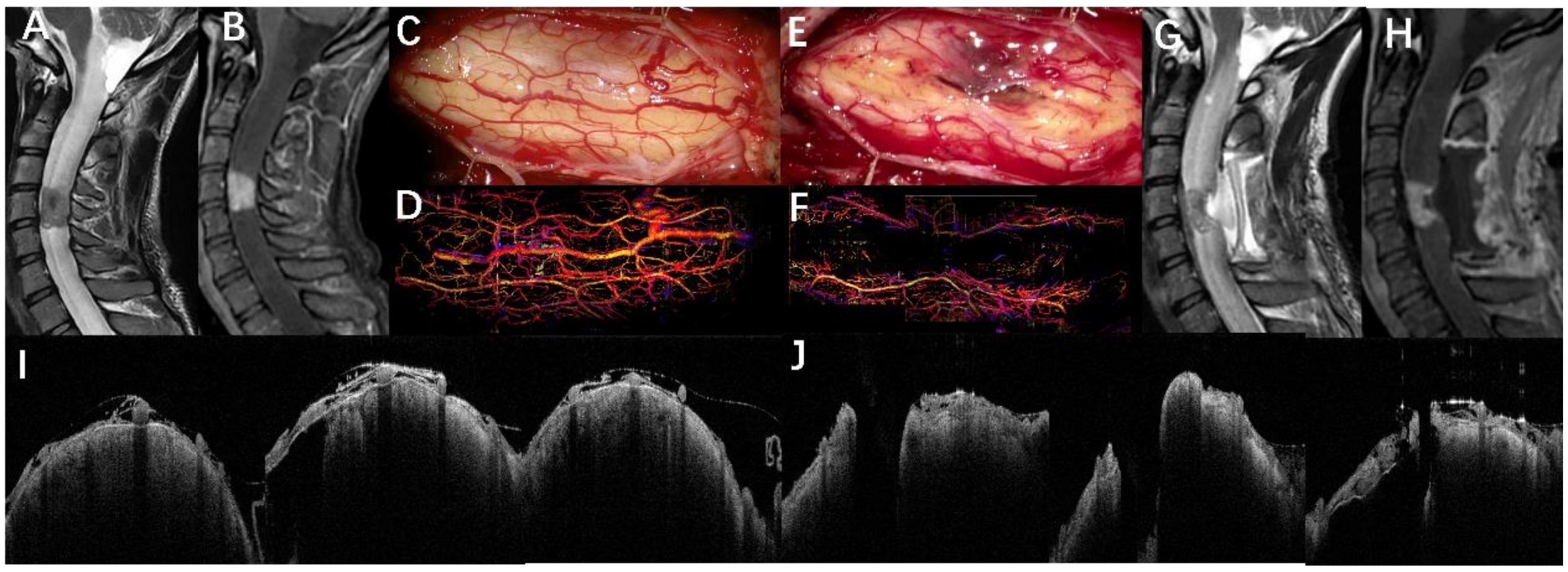

**SFig.4 A case of subependymal tumor:** (A-B) Preoperative MRI shows the tumor located within the cervical spinal canal, with a clearly enhanced lesion that has broken through the spinal cord surface. (C) Intraoperative macroscopic view of the tumor shows the tumor surface as light grayish-red, located on the right side of the tumor. The surface vasculature appears sparse and not significantly disorganized. (D) OCTA vascular imaging shows sparse vasculature with no apparent disorganization, with some surface vessels corresponding to the macroscopic view in panel C. (E) Macroscopic view of the surgical cavity after tumor resection and pia mater suturing. (F) OCTA vascular image after tumor resection and pia mater suturing shows moderate vascular density on the spinal cord surface, with no vascular signals inside the surgical area, corresponding to the macroscopic view in panel E. (G-H) Postoperative MRI shows residual tumor following resection, with no hematoma or fluid accumulation in the surgical area. (I) Selected OCT B-scan structural images from the rostral to caudal ends of the tumor show heterogeneous signals within the tumor, with calcification and cystic degeneration signals present. (J) Selected OCT B-scan structural images post-tumor resection show gaps after pia mater suturing, with optical signals similar to those observed before tumor resection, suggesting residual tumor, corresponding to panels G-H.

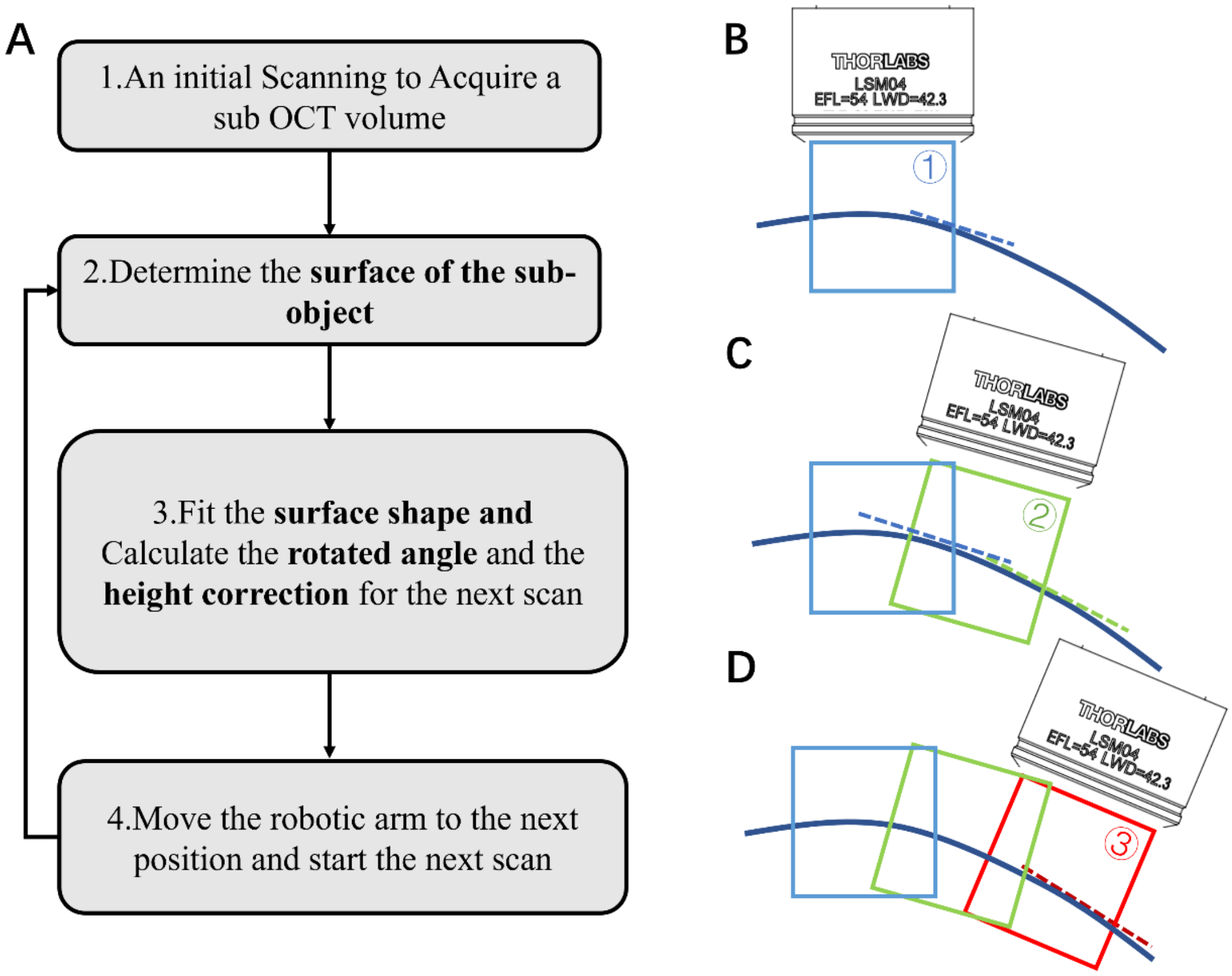

**SFig. 5 Automatic large-scale scanning mechanism employed by the FACT-ROCT system:** (A) Flowchart of the automatic large-scale scanning process. The core component involves the movement of the robotic arm driven by the previous OCT volume data to achieve extensive scanning. (B)-(D) Schematic diagrams illustrating three consecutive OCT volume scans.

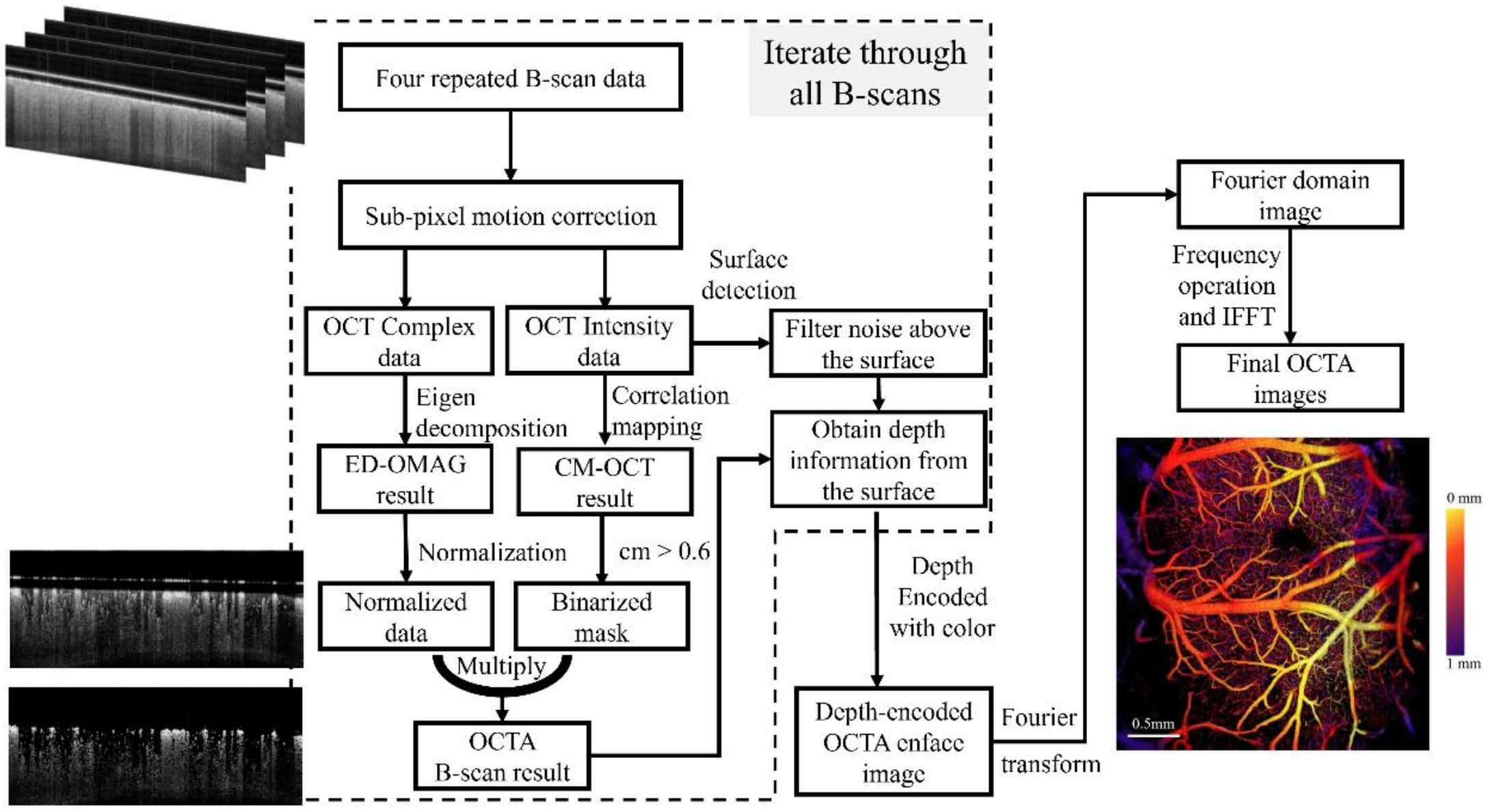

**SFig. 6 The flow chart of OCTA algorithm:** The process begins with four repeated B-scan data, followed by sub-pixel motion correction and surface detection. OCT complex data undergo eigen decomposition, yielding eigen decomposition optical microangiography (ED-OMAG) results, while intensity data are processed through correlation mapping to generate correlation mapping OCT (CM-OCT) results. After normalization and binarization, the OCTA B-scan results are obtained. Surface depth information is retrieved and encoded with color. The depth-encoded OCTA enface image is then subjected to a Fourier transform, resulting in the final OCTA images through frequency operation and inverse Fourier transform (IFFT). The final image on the right shows a depth-encoded vascular map in an enface OCTA image, with color representing different depths.

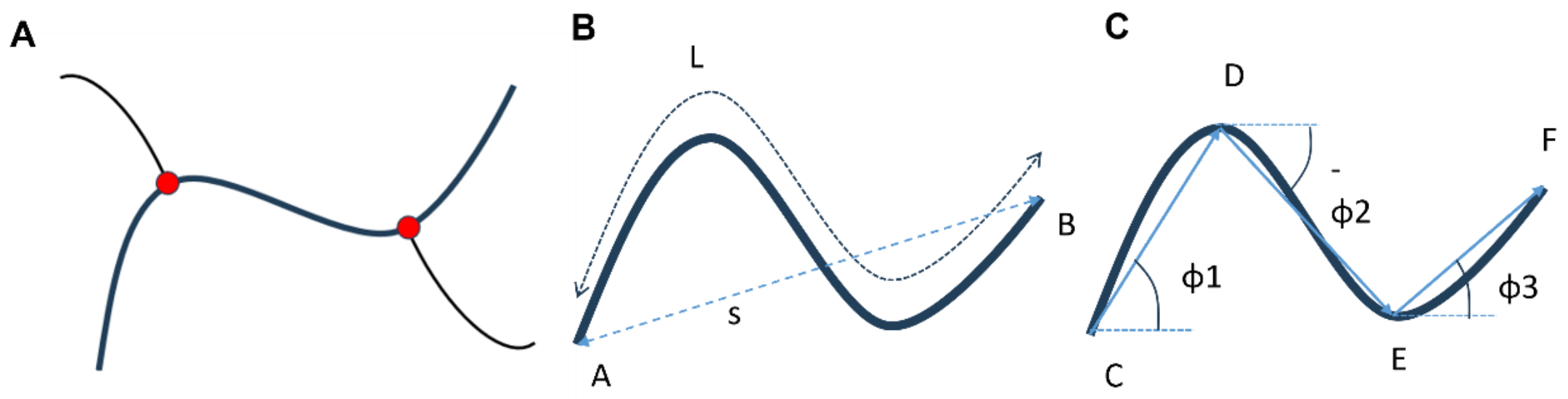

**SFig. 7 Illustration of vascular parameter analysis calculations:** (A) Vascular branch points indicated by red dots. (B) The definition of vascular tortuosity for a single vessel. (C) The orientation azimuth statistics represent the trajectory variability of a single vessel.

**Table S1: The table of patient cohort**

| # | Age at surgery | Sex | Diagnosis | Recurrence(Y/N) | Tumor location |
|---|---|---|---|---|---|
| **1** | 11 | F | Ganglioglioma (WHO1) | N | Medulla oblongata - C3 |
| **2** | 33 | F | Subependymoma (WHO1) | N | C4-C5 |
| **3** | 13 | M | Pilocytic myxoid astrocytoma (WHO2) | Y | Medulla oblongata - T10 |
| **4** | 52 | M | Diffuse astrocytoma (WHO2) | N | C2-C4 |
| **6** | 29 | F | Ependymoma (WHO2) | N | Medulla oblongata - C2 |
| **7** | 33 | M | Ependymoma (WHO2) | N | C6-T1 |
| **8** | 26 | M | Anaplastic astrocytoma (WHO3) | N | T4-T7 |
| **9** | 4 | M | Anaplastic ependymoma (WHO3) | Y | C1-C7 |
| **10** | 18 | F | Anaplastic ganglioglioma (WHO3) | N | Medulla oblongata - C3 |
| **11** | 54 | M | Diffuse midline glioma (WHO4) | Y | C2-C6 |
| **12** | 56 | M | Diffuse midline glioma (WHO4) | N | T11-T12 |
| **13** | 40 | F | Diffuse midline glioma (WHO4) | N | T11-T12 |
| **14** | 13 | M | Diffuse midline glioma (WHO4) | N | T4-T7 |
| **15** | 55 | F | Angioreticuloma | N | Medulla oblongata |
| **16** | 41 | F | Teratoma | Y | T11-L2 |
| **17** | 9 | F | Myelitis | N | C3-C4 |
| **18** | 41 | M | Epidermoid cyst | N | T12-L1 |
| **19** | 21 | F | Metastatic malignant meningioma | Y | T11-T12 |
| **20** | 10 | F | Ewing sarcoma | N | Medulla oblongata - C2 |
| **21** | 71 | M | Spinal arteriovenous fistula | N | T9-T10 |